\documentclass{jfm}
\usepackage{graphicx}
\usepackage{epstopdf, epsfig}
\usepackage{siunitx}
\usepackage{physics}
\usepackage{multirow}   % multirow cells in tables

\usepackage{natbib}
\usepackage{amsbsy}
\usepackage{psfrag}
\usepackage{amsmath}
\usepackage{bm}
\usepackage{soul}
\usepackage{verbatim}
\usepackage[dvipsnames]{xcolor}
\usepackage{url}
\usepackage[colorlinks=true,
             linkcolor=blue,
             urlcolor=blue,
             citecolor=blue]{hyperref}
\usepackage{lineno}  % line numbers
\usepackage[noabbrev, nameinlink]{cleveref}
\crefformat{equation}{(#2#1#3)}  % matches \eqref
\crefrangeformat{equation}{(#3#1#4) to~(#5#2#6)}

\newcommand{\IEA}{IEA \SI{15}{\mega\watt} }
\newcommand{\winddir}{\psi}
\newcommand{\BV}{Brunt-V\"ais\"al\"a }
\newcommand*{\ie}{i.e.,\ }
\shorttitle{Coriolis effects on wind turbine wakes in the ABL}
\shortauthor{K. S. Heck and M. F. Howland}

\title{Coriolis effects on wind turbine wakes across atmospheric boundary layer regimes}

\author{Kirby S. Heck\aff{1} 
  \and Michael F. Howland\aff{1} 
    \corresp{\email{mhowland@mit.edu}}
    }

\affiliation{\aff{1}Department of Civil and Environmental Engineering, Massachusetts Institute of Technology,
Cambridge, MA 02139, USA}

\begin{document}

\maketitle

\begin{abstract}
% This file contains instructions for authors planning to submit a paper to the {\it Journal of Fluid Mechanics}. These instructions were generated in {\LaTeX} using the JFM style, so the {\LaTeX} source file can be used as a template for submissions. The present paragraph appears in the \verb}abstract} environment. All papers should feature a single-paragraph abstract of no more than 250 words, which provides a summary of the main aims and results. 
Wind turbines operate in the atmospheric boundary layer (ABL), where Coriolis effects are present. 
As wind turbines with larger rotor diameters are deployed, the wake structures that they create in the ABL also increase in length. 
Contemporary utility-scale wind turbines operate at rotor diameter-based Rossby numbers, the nondimensional ratio between inertial and Coriolis forces, of $\mathcal{O}(100)$ where Coriolis effects become increasingly relevant.
Coriolis forces provide a direct forcing on the wake, but also affect the ABL base flow, which indirectly influences wake evolution.
These effects may constructively or destructively interfere because both the magnitude and sign of the direct and indirect Coriolis effects depend on the Rossby number, turbulence, and buoyancy effects in the ABL.
Using large eddy simulations, we investigate wake evolution over a wide range of Rossby numbers relevant to offshore wind turbines. 
Through an analysis of the streamwise and lateral momentum budgets, we show that Coriolis effects have a small impact on the wake recovery rate, but Coriolis effects induce significant wake deflections which can be parsed into two regimes. 
For high Rossby numbers (weak Coriolis forcing), wakes deflect clockwise in the northern hemisphere. 
By contrast, for low Rossby numbers (strong Coriolis forcing), wakes deflect anti-clockwise. 
Decreasing the Rossby number results in increasingly anti-clockwise wake deflections.
The transition point between clockwise and anti-clockwise deflection depends on the direct Coriolis forcing, pressure gradients, and turbulent fluxes in the wake. 
At a Rossby number of 125, Coriolis deflections are comparable to wake deflections induced by $\SI{17}{\degree}$ of yaw-misalignment. 
\end{abstract}

% \begin{keywords}
% Authors should not enter keywords on the manuscript, as these must be chosen by the author during the online submission process and will then be added during the typesetting process (see http://journals.cambridge.org/data/\linebreak[3]relatedlink/jfm-\linebreak[3]keywords.pdf for the full list)
% \end{keywords}

\section{Introduction}
\label{sec:intro}

% - Introduce the field.
	% - Introduce the problem.
	% - What is the knowledge gap?
	% - What are you doing about it?
	% - How are you doing it?
	% - Why is it important?

% Maybe 1 paragraph on even larger picture: 
Wind energy is a necessary component for rapid decarbonization of the global electricity sector \citep{veers_grand_2019}. 
Ambitious climate goals to cut greenhouse gas emissions have led to recent policies propelling the expansion of wind energy development worldwide, particularly for offshore wind \citep{iea_renewables_2022}. 
In parallel, the rated power, hub-height, and rotor diameter of offshore wind turbines are projected to continue to increase in the foreseeable future \citep{diaz_review_2020}. 
% 1 more connecting sentence? 

Wind turbines, which extract power from the incoming wind, produce momentum and mean kinetic energy deficient regions downwind called wakes. 
Wind turbine wakes, which are large, streamwise flow structures stretching over 10 turbine diameters ($D$) in length \citep{hogstrom_field_1988}, adversely impact downwind turbines. 
Lower wind speeds within turbine wakes decrease the power production of waked turbines \citep{barthelmie_modelling_2007}. 
For offshore wind farms, wake interactions can reduce annual energy production 10-20\% \citep{barthelmie_quantifying_2010}. 
Mitigating wake interactions between turbines either through wind farm design or control relies on accurate predictions of wake evolution in the presence of atmospheric boundary layer (ABL) physics \citep{meyers_wind_2022}. 

All utility scale wind turbines operate in the ABL, where many physical processes can influence wake evolution, including Coriolis effects.  
Coriolis forces---which are present in the ABL due to the rotation of the earth---redirect winds and redistribute kinetic energy in the atmosphere \citep{stull_introduction_1988}. 
The dynamical importance of Coriolis forces in a rotating flow is described by the Rossby number $Ro = U_c/(\omega_c L_c)$, where $U_c$ is a characteristic velocity scale, $\omega_c$ is a characteristic angular velocity, and $L_c$ is a characteristic length scale of the flow.  
The Rossby number represents a ratio of inertial forces to Coriolis forces. 
%Large Rossby numbers ($Ro \gg 1$) imply that fluid inertia ($\sim U_c^2 / L_c$) is significantly more important to the dynamics of the system than Coriolis forces ($\sim \omega_c U_c$) 
% because the characteristic flow velocity $U_c$ is large relative to the velocity associated with rotation $\omega_c L_c$
%, and vice versa for small Rossby numbers ($Ro \sim 1$). 
Thus, the relative strength of Coriolis forces increases with increasing length scales of the flow. 
As wind turbines and, correspondingly, their wakes increase in size, the influence of Coriolis forces on wake evolution, recovery, and deflection may differ from the previous generation of wind turbines that have been studied in existing literature. 
Furthermore, because wind turbines operate at a range of inflow wind speeds, the relative importance of Coriolis forces on the wake dynamics of a turbine will change across regions of turbine operation as a result of changing $U_c$. 
Therefore, it is important to study the effects that Coriolis forces have on wind turbine wake evolution for a range of Rossby numbers. 

In the ABL, the characteristic velocity is the geostrophic wind speed $G$ and the angular velocity is the Coriolis parameter $f_c = 2\omega \sin\phi$, where $\omega = \SI{7.29e-5}{\radian\per\second}$ is the rotation rate of the earth and $\phi$ is the latitude. 
The characteristic length scale for wind turbine wake dynamics is the wind turbine diameter $D$ \citep{meyers_optimal_2012}. 
As the turbine diameter increases, the relative importance of Coriolis forces increases ($Ro$ decreases). 
Note that while the turbine diameter is chosen for the characteristic length scale, wind turbine wakes are typically an order of magnitude larger than the turbine diameter \citep{hogstrom_field_1988}. 
Still, the turbine diameter is selected as the characteristic length scale in the Rossby number because the diameter of a turbine is fixed, while the wake length will depend on properties of the ABL, such as inflow turbulence. 

The presence of Coriolis forces in the ABL has two primary effects on wake development \citep{van_der_laan_why_2017}. 
First, because wind turbine wakes are regions of lower velocity than the surrounding wind flow, the direct Coriolis forcing, which is proportional to the velocity, is different between the wake and the surrounding flow \citep{howland_influence_2018}. 
Second, Coriolis forces alter the structure of the background ABL by affecting the vertical shear of wind speed and direction (gradients of wind speed and direction, respectively) \citep{wyngaard_turbulence_2010}. 
This background ABL wind, affected by Coriolis forces, flows into the turbine and also influences wake dynamics. 

In previous literature, Coriolis forces have been show to alter the dynamics of wakes of individual wind turbines and of wind farms. 
Using large eddy simulations (LES), \citet{abkar_influence_2016} studied the wake dynamics of a free standing wind turbine in the ABL subjected to Coriolis forces (geostrophic pressure gradient forcing) compared with the wake dynamics in a turbulent boundary layer without Coriolis forcing (pressure driven boundary layer). 
They found that the presence of Coriolis forces enhances turbulence kinetic energy production driven by wind direction shear that is only present in the geostrophic pressure gradient driven boundary layer, which results in faster wake recovery than in the pressure driven boundary layer without Coriolis forcing.  % check if churchfield (2018) study also finds this
Wind direction shear, caused by Coriolis forcing, also creates a skewed wake structure because different vertical levels of the wind turbine rotor are subjected to different inflow wind directions. 
In addition to numerical simulations, the skewed wake structure has been observed in field experiments \citep{magnusson_influence_1994, bodini_three-dimensional_2017}.  % cite others? 
In summary, the presence of Coriolis forces in the ABL has been shown to alter the recovery and structure of wind turbine wakes relative to wakes in pressure-driven boundary layers. 

In addition to changes in the wake recovery and structure, Coriolis effects have been shown to affect wake deflection. 
In Reynolds-averaged Navier--Stokes (RANS) simulations of a wind farm wake at a diameter-based Rossby number of $Ro = 952$, \citet{van_der_laan_why_2017} observed clockwise wake deflection (as viewed from above) as a result of Coriolis effects in the Northern hemisphere. 
They also observed clockwise wake deflection in RANS simulations of an isolated wind turbine wake. 
The clockwise wake deflection was attributed to the vertical turbulent entrainment of lateral momentum in the wake recovery process, which was shown to add clockwise-turning flow from aloft into the wake region. 
Winds typically turn clockwise with ascending height in the northern hemisphere due to the Ekman spiral \citep{ekman_influence_1905}, which is caused by Coriolis effects on wind shear in the ABL. 
In contrast, the direct Coriolis forcing in the wind turbine wake region turns lower velocity wakes anti-clockwise in the ABL \citep{van_der_laan_why_2017, howland_influence_2018}. 
Therefore, for wakes generated in flat, horizontally homogeneous terrain, the vertical entrainment and direct Coriolis forcing mechanisms can be in opposition. 
The imbalance of the turbulent flux of clockwise-turning momentum over the anti-clockwise direct Coriolis forcing in a wake at $Ro = 952$ has been used to explain the clockwise wake deflections resulting from Coriolis effects \citep{van_der_laan_why_2017}. 

However, differing results for the magnitude and direction of wake deflection due to Coriolis effects in the ABL have been reported in existing literature. 
In LES of a five-row turbine array at $Ro = 1005$,  \citet{nouri_coriolis_2020} observed a slight clockwise wake deflection of approximately 0.2$D$ at a distance of 6$D$ after the first turbine row due to Coriolis effects in the ABL. 
An investigation of wakes in a stably stratified boundary layers by \citet{englberger_does_2020} at $Ro = 1000$ found that wakes deflect clockwise at hub height, regardless of the rotation direction of the turbine rotor. 
Additionally, \citet{qian_control-oriented_2022} compared lidar and LES data of an offshore wind turbine wake in the ABL at $Ro = 1013$ and observed a similar amount of clockwise wake deflection as \citet{nouri_coriolis_2020}. 
These results agree with the clockwise wake deflection observed by \citet{van_der_laan_why_2017}. 
In contrast, other studies have observed zero or anti-clockwise turning in turbine wakes due to Coriolis effects. 
In LES of a neutrally stratified and stably stratified ABL at $Ro = 574$, \citet{gadde_effect_2019} observed an initial anti-clockwise wake deflection of wakes within a wind farm, which transitioned to clockwise deflection after the first 2-3 rows of turbines. 
They noted that the transition to clockwise deflection was related to the vertical turbulent momentum flux into the wake region. 
Negligible wake deflection was reported in single-turbine studies using RANS \citep{van_der_laan_predicting_2015} and LES \citep{abkar_analytical_2018, mohammadi_curled-skewed_2022} for Rossby numbers between $550$ and $1200$. 
Finally, anti-clockwise wake turning was observed in LES of offshore wind farms by \citet{dorenkamper_impact_2015} in land-sea transition at $Ro = 815$ and by \citet{allaerts_boundary-layer_2017} in neutrally stratified conditions at $Ro = 1200$. 

Several challenges are highlighted by the spread of wake deflection results in previous literature. 
For example, insufficient time averaging can conflate wake deflection results with unsteady ABL turbulence, as noted by \citep{churchfield_using_2016}. 
Additionally, wind speed and direction shear depend on the strength of Coriolis forcing relative to the strength of inertial forces in the ABL (throughout this paper, relative Coriolis forcing strength refers to the comparison between inertial forces and Coriolis forces, as quantified by the Rossby number). 
Because neutral boundary layers are sensitive to their heating history \citep{tennekes_model_1973, allaerts_boundary-layer_2017}, wind shear and turbulence, which affect the wake deflection, can vary significantly for the same Rossby number. 
For stable boundary layers, wind shear and turbulence also vary based on the surface cooling rate, which is independent of the Rossby number. 
Unless these factors are controlled, comparisons of wake dynamics in ABL conditions may have contradictory results. 

\begin{figure}
    \centering
    \includegraphics[width=1\linewidth]{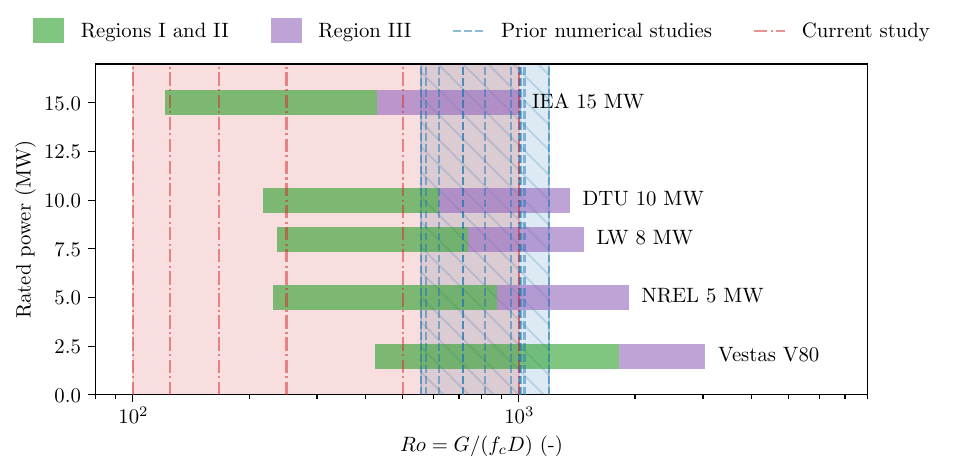}
    \caption{Relevant Rossby numbers for several wind turbines models in the mid-latitudes ($\phi = 45^\circ$). Green bars indicate Region I and II operation where turbines aim to maximize the power they extract from the wind, while purple bars indicate Region III operation where individual turbines curtail power generation at high wind speeds. Previous numerical experiments are marked with blue dashed lines, and simulations from the current study are shown as a red dash-dotted lines.}
    \label{fig:ro_intro}
\end{figure}

Previous literature has also focused on a relatively narrow range of Rossby numbers between $Ro = 550$ and $Ro = 1200$. 
For the \SI{126}{\meter} rotor diameter NREL~5~MW turbine \citep{jonkman_definition_2009} at mid-latitudes, this represents a wind speed range of $6.5-13$\si{\meter\per\second}. 
However, trends to both manufacture larger turbines (increasing $D$) as well as install new wind turbines in regions of lower average wind speed (decreasing $G$) push turbines into lower Rossby number regimes. 
For example, the \IEA reference turbine \citep{gaertner_definition_2020} has a \SI{240}{\meter} rotor diameter, nearly twice as large as the NREL~5~MW. 
In contrast to the NREL~5~MW turbine, which produces rated power at $Ro = 880$ in the mid-latitudes, the \IEA reference turbine rated wind speed corresponds to $Ro = 430$ where the ratio of Coriolis to inertial forces is twice as large. 
Meanwhile, the \IEA turbine cut-in occurs at $Ro = 120$.
\Cref{fig:ro_intro} shows this trend for larger offshore wind turbines to operate in lower Rossby number regimes---which will be investigated in this study---overlaid on a survey of prior numerical studies on turbine wakes including Coriolis forces. 

% What is the knowledge gap: 
    % Where are the regimes when (direct) Coriolis forcing dominate? What about regimes of (indirect) Coriolis effects? 
    % How do we study this problem independently? 
    % Why are there differing results reported by different studies? 
    % How does Coriolis affect wake recovery, deflection, and evolution parametrically? 

% What are you doing about it? 
    % We are investigating wake dynamics of controlled experiments parametrically and for a wider range of Ro than currently reported in the literature

% How are we doing it? 
    % By altering the wind speed in these simulations, we can change the Rossby number and therefore vary the strenght of Coriolis forcing. 
    % We use LES to be able to tightly control each problem setup

% Why is this important? 
    % Implications for next generation turbines: wake losses, wake steering, wind farm control. There has been an increasing focus on wind farm flow control (cite Meyers et al. (2022)) to increase collective power production of wind farms. This relies on accurate wake modeling which typically does not include effects of Coriolis forcing, wind shear, etc. 

The existing body of literature addressing Coriolis effects on wake evolution focuses on Rossby numbers pertinent to the previous generation of wind turbines. 
Larger turbines will operate in Rossby number regimes where ABL dynamics may differ from the Rossby number regimes at which present-day wind turbines operate, but the demarcation of different regimes is not clear from the existing literature. 
% Furthermore, the influence of Coriolis effects on wind turbine wake dynamics 
In this study, we explore a wide range of Rossby numbers relevant for flows at the contemporary commercial turbine scale. 
In particular, we consider the range of Rossby numbers relevant to the \IEA reference turbine. 
For this turbine in the mid-latitudes ($\phi = 45^\circ$), Rossby numbers between 100 and 500 span the range of wind speeds in regions I and II of wind turbine control---where wake interactions are most relevant---as shown in \cref{fig:ro_intro}. 

Here, we present a suite of large eddy simulations of increasing complexity which parametrically vary the relative strength of Coriolis forcing. 
Three types of inflow are considered. 
First, wake dynamics are investigated in turbulence-free, shear-free uniform inflow to isolate the direct Coriolis forcing on the wake from the Coriolis effects on the wind shear. 
Then, wind turbine wakes are simulated in neutrally stratified boundary layers to study both the direct Coriolis forcing and Coriolis effects on wind shear which influence wake dynamics. 
We focus on neutral boundary layers in this work to limit the problem complexity and to isolate the competing direct and indirect effects of Coriolis forcing on wake evolution. 
Finally, we simulate wakes in conventionally neutral conditions, where the neutral boundary layer is capped by a stable free atmosphere. 
In each of these inflow conditions, we examine the primary momentum transport mechanisms governing wake recovery, deflection, and structure as a function of the Rossby number. 
The goal is to understand how Coriolis effects influence wake evolution across the range of Rossby numbers that will be encountered by existing and future wind turbines.

% last paragraph: roadmap
The remainder of this study is organized as follows. 
In \cref{sec:LES}, we introduce the LES numerical setup. 
Following this in \cref{sec:method_budgets}, we present a streamtube-based methodology for analyzing the time-averaged wake recovery, deflection, and dynamics. 
Results are shown in \cref{sec:results} for the inflow conditions, wake structure and evolution, and momentum budget analysis, followed by concluding remarks and discussion of future work in \cref{sec:conclusion}.

\section{LES numerical setup}
\label{sec:LES}

In this study, large eddy simulations are used to solve the filtered, incompressible Navier--Stokes equations under the infinite Reynolds number limit with the Boussinesq approximation for buoyancy,
\begin{gather}
    \label{eq:continuity}
    \pdv{u_i}{x_i} = 0,   \\
    \label{eq:NSE_LES}
    \pdv{u_i}{t} + u_j\pdv{u_i}{x_j} = 
    - \pdv{p}{x_i} 
    + \frac{\delta_{i3}}{\theta_0 Fr^2} (\theta - \theta_0)
    - \pdv{\tau_{ij}}{x_j} 
    - \frac{1}{Ro}\varepsilon_{ij3} (G_j - u_j)
    % - \pdv{p^G}{x_i} 
    + f_{t, i}, 
\end{gather}
where $u_i$ is the filtered non-dimensional velocity in the $x_i$ direction, $t$ is non-dimensional time, $p$ is a non-dimensional pressure, and $\theta$ is potential temperature. 
The $i = 1, 2$, and $3$ indices correspond with the streamwise ($x$), lateral ($y$), and vertical ($z$) directions, respectively. 
The reference potential temperature is $\theta_0$, $\delta_{ij}$ is the Kronecker delta, and $\varepsilon_{ijk}$ is the permutation operator. 
Note that the geostrophic balance has been substituted such that the ($1/Ro$) $\varepsilon_{ij3} G_j$ term represents the geostrophic pressure gradient, and $G_j$ is the geostrophic velocity vector. 
The non-dimensional turbine forcing is given by $f_{t, i}$, and subgrid stresses are given by $\tau_{ij}$. 
The Froude number $Fr = G/\sqrt{gD}$ governs the magnitude of buoyancy forces, where $g$ is the gravitational acceleration. 
Following the definition of the Rossby number in \cref{sec:intro}, the velocity scale is $G$ and the characteristic length scale is the turbine diameter $D$. 
% The Rossby number based on the turbine diameter is defined as $Ro = G/(f_c D)$ where $f_c = 2\Omega \sin\phi$ is the Coriolis parameter, $\Omega = \SI{7.29e-5}{\radian\per\second}$ is the rotation rate of the earth, and $\phi$ is latitude. 
The traditional approximation of Coriolis forces is enforced to focus this study on the vertical component of Earth's rotation. 
The traditional approximation neglects the horizontal component of Earth's rotation, which affects the ABL structure with and without the presence of wind turbines \citep{howland_influence_2020}. 
However, the focus of this study is strictly on the influence of the vertical component of Earth's rotation on wind turbine wakes. 
The prognostic equation for the filtered potential temperature $\theta$ is given by 
\begin{equation}
    \label{eq:temp_eqn}
    \pdv{\theta}{t} + u_j\pdv{\theta}{x_j} = \pdv{q_{ij}}{x_j}, 
\end{equation}
where $q_{ij}$ is the subgrid-scale heat flux. Scalar diffusivity is modeled with a Prandtl number $Pr = 0.4$ \citep{howland_influence_2020}. 

The incompressible flow solver Pad\'{e}Ops\footnote{\url{https://github.com/Howland-Lab/PadeOps}} \citep{howland_influence_2020, ghate_subfilter-scale_2017} is used to solve the filtered, incompressible Navier--Stokes equations. 
Fourier collocation is used in the horizontal directions while a sixth-order staggered compact finite difference scheme is used in the wall-normal direction \citep{nagarajan_robust_2003}. 
A fourth-order strong stability preserving variant of Runge-Kutta scheme is used for time integration \citep{gottlieb_strong_2011} and the sigma subgrid scale model \citep{nicoud_using_2011} is used to model the subgrid stresses $\tau_{ij}$. 
Periodic boundary conditions are used in the streamwise and lateral directions, and a fringe region \citep{nordstrom_fringe_1999} is used in the final 25\% of the domain in the streamwise direction to replenish momentum lost in the wake region. 

The wind turbine is implemented as an actuator disk model (ADM) introduced by \citet{calaf_large_2010} and further developed by \citet{shapiro_filtered_2019}. 
The ADM exerts a force normal to the disk face that is proportional to the disk-averaged velocity $u_d$. 
The magnitude of the thrust force is computed $F_T = (1/2) \rho A_d u_d^2 C_T'$, where $\rho$ is air density, $A_d = \pi D^2/4$ is the area of the ADM, and $C_T'$ is the thrust coefficient of the turbine \citep{calaf_large_2010}. 
A local thrust coefficient $C_T' = 1.33$ is used in all simulations, which corresponds to $C_T = 3/4$ following one-dimensional momentum theory. 
The thrust coefficient is held fixed, and not modified as a function of the incident wind speed in this study.
The forcing is distributed over a Gaussian smoothing kernel $\mathcal{R}(x, y, z)$ given by \citet{shapiro_filtered_2019} to reduce numerical oscillations in the flow fields. 
For all simulations, the smoothing filter width $\Delta = 2.5h$ is used, where $h \equiv (\Delta x^2 + \Delta y^2+\Delta z^2)^{1/2}$ and $\Delta x_i$ is the grid spacing in the $x_i$ direction. 
The thrust correction factor derived by \citet{shapiro_filtered_2019} is not used in these simulations. 

The turbine size is based off of the IEA~\SI{15}{\mega\watt} reference turbine \citep{gaertner_definition_2020}. 
The diameter of the reference turbine is $D = \SI{240}{\meter}$ and the turbine hub height is $z_h = \SI{150}{\meter}$. 
The range of rotor diameter-based Rossby numbers that span the operational range of the \IEA reference turbine in the mid-latitudes ($\phi = 45^\circ$) are $Ro = 120$ at the cut-in speed of \SI{3}{\meter\per\second} and $Ro = 430$ at the rated wind speed of \SI{10.6}{\meter\per\second}. 
To vary the relative strength of Coriolis forcing, the Rossby number is changed for each independent simulation. 
We sweep over a range of Rossby numbers by varying $Ro^{-1}$ linearly from 0.002 ($Ro = 500$) to 0.010 ($Ro = 100$). 
This parameter range overlaps with region II operation of the \IEA reference turbine \citep{gaertner_definition_2020} as shown in \cref{fig:ro_intro}. 
Additionally, simulations at $Ro = 1000$ are included to compare with existing literature. 
We note that because of the non-dimensionalization of the Coriolis forcing, the Rossby numbers studied here can be mapped to different wind speed ranges of other turbine diameters or at different latitudes. 

All simulations are performed with a computational domain of length $L_x = 38.4D$ (\SI{9.2}{\kilo\meter}) in the streamwise direction and a cross-section $L_y \times L_z$ of $12.8D \times 12.8D$ (\SI{3.1}{\kilo\meter} $\times$ \SI{3.1}{\kilo\meter}) unless noted otherwise. 
The domain is rectangular and uses an evenly spaced, uniform grid of $384 \times 256 \times 256$ points, resulting in 19 points in the vertical direction across the ADM, which is sufficient resolution for convergence of turbine forcing \citep{stevens_comparison_2018}. 
The ADM is placed $5D$ from the inlet of the computational domain. 
Simulations are run until time-averaged statistics within the wind turbine wake have converged, which varies based on the dynamics of the problem setup. 
Three different problem setups of increasing complexity are described in \cref{ssec:setup_uniform} and \cref{ssec:setup_ABL} to study the influence of Coriolis forces on wind turbine wakes.

\subsection{Uniform inflow simulation setup}
\label{ssec:setup_uniform}

% The most elementary experiment for studying the effects of the Coriolis force on a wind turbine wake is simulating uniform, turbulence-free inflow. 
A wind turbine wake in uniform, turbulence-free inflow serves as the canonical reference case to study interactions between wakes and Coriolis effects.  
In uniform inflow simulations, transport mechanisms related to Coriolis forces can be isolated without confounding variables such as freestream turbulence, wind shear, or thermal stratification. 
A single ADM turbine is centered laterally and vertically in the computational domain, and the bottom and top domain boundaries are slip walls. 
The velocity is initialized to the geostrophic wind velocity in the $x$ direction in the entire domain, and the geostrophic wind speed is set based on the Rossby number $Ro$. 
There is no velocity shear, no thermal stratification, and zero turbulence in the prescribed inflow. 
The prognostic equation for the potential temperature is omitted in the uniform inflow simulations. 
Time-averaged statistics are taken over an interval of eight flow-through times ($8 L_x/G$), which is sufficient for turbulence free, uniform inflow simulations \citep{howland_wake_2016}. 
An initial transient period of $2 L_x/G$ allows a buffer time for the wake to develop before averaging begins.

\subsection{Atmospheric boundary layer simulation setup}
\label{ssec:setup_ABL}

Utility-scale turbines operate in the ABL, where wind speed and direction shear is present over the rotor area \citep{stevens_flow_2017}. 
Wind shear in the ABL results from the balance of the driving geostrophic pressure gradient, Coriolis forcing, and friction. 
Therefore, the magnitude of wind speed and direction shear in the ABL depends on the Coriolis forcing, here quantified using the Rossby number. 
Wind speed and direction shear both affect wake dynamics \cite[\eg][]{abkar_influence_2016}. 
Therefore, in the ABL, Coriolis forcing has a direct effect on wake evolution, as in the uniform case, but also has indirect effects that manifest through the wind shear. 
As the Rossby number is changed in the ABL, the wind speed and direction shear will change in addition to the direct Coriolis forcing.

Buoyancy effects modify wind shear and turbulent fluxes in the ABL \citep{stull_introduction_1988}. 
Thermal stratification, which may be stable, neutral, or unstable (convective), also alters wake evolution \citep{abkar_influence_2015, xie_numerical_2017}. 
We simulate neutrally stratified ABLs, which have no vertical potential temperature gradient in the boundary layer region, to study the effects of Coriolis forcing on wake development while minimizing additional complexity due to buoyancy effects. 
Two types of neutrally stratified ABLs are studied. 
The truly neutral boundary layer (TNBL), also known as the turbulent Ekman layer \citep{ekman_influence_1905}, is neutrally stratified throughout the vertical domain. 
That is, the TNBL is isothermal. 
Initialization for the TNBL is described in \cref{sssec:TNBL_init}. 
While no atmospheric boundary layers are truly neutral due to the presence of stable stratification in the free atmosphere \citep{stull_introduction_1988}, parametrically analyzing the influence of Coriolis forcing on wind turbine wakes in the TNBL adds the complexity of wind speed and direction shear without considering the effects of thermal stratification. 
We also simulate conventionally neutral boundary layer (CNBL) inflow \citep{zilitinkevich_further_2007}. 
The CNBL adds a stably stratified free atmosphere above a neutrally stratified, turbulent boundary layer, which is more similar to neutral ABLs observed in the environment \citep{rampanelli_method_2004}. 
The initialization for the CNBL potential temperature field is given in \cref{sssec:CNBL_init}. 

After initializing the potential temperature field, pseudo-random potential temperature perturbations are added to the bottom \SI{100}{\meter} of the ABL to spin up turbulence. 
The initial velocity profile is uniform and geostrophic, aligned in the $x$-direction. 
The bottom boundary condition uses a Monin-Obukhov wall model \citep{moeng_large-eddy-simulation_1984} with a surface roughness $z_0 = \SI{0.1}{\milli\meter}$, which is characteristic of offshore wind conditions \citep{allaerts_boundary-layer_2017, stull_introduction_1988}. 
The heat flux at the ground is set to zero to enforce neutrally stratified conditions in the ABL. 
A Rayleigh damping region \citep{allaerts_boundary-layer_2017} is used in the upper 25\% of the vertical domain to absorb perturbations propagated into the free atmosphere. 
The spin-up simulation is run until $t = 15/f_c \approx \SI{40}{\hour}$ to allow inertial oscillations to decay such that the ABL reaches statistical quasi-equilibrium. 

Before placing a turbine in the ABL, the flow at hub height is aligned to the $x$ direction by rotating the domain in the spin-up simulation with a wind angle controller which imparts a pseudo-Coriolis force \citep{sescu_control_2014, howland_influence_2020}. 
The rotation only aligns the flow with the computational domain and  does not affect the ABL structure or statistics \citep{sescu_control_2014, howland_influence_2018}. 
Once the wind angle is rotated and the boundary layer reaches a new equilibrium, the wind angle controller is turned off. 
Because the ABL is in quasi-equilibrium at the end of the spin-up simulation, the wind angle drift is small, and the hub-height wind angle remains within $\pm 0.5^\circ$ of the original direction for all simulations after the wind angle controller is removed. 

The concurrent-precursor method \citep{stevens_concurrent_2014} is used to simulate a finite wind turbine wake. 
An ADM turbine is placed in the boundary layer at a hub height of $z_h = \SI{150}{\meter}$ and centered laterally in the domain of the primary simulation. 
The ADM does not use a yaw angle controller because the wind angle drift is small, so the wind turbine remains yaw-aligned throughout the simulation. 
As a result, the turbine forcing remains exclusively in the $x$-direction. 
A fringe region is active for the primary simulation only, which restores the flow to the state of the precursor simulation. 
The concurrent simulations are time-averaged over one inertial period $T = 2\pi/f_c \approx \SI{17}{\hour}$ to allow wake statistics to converge, and to avoid averaging over a partial period of inertial oscillations. 
This is a sufficiently long time-averaging window as shown in \cref{appx:time_averaging}. 
A transient period of $2 L_x/G$ is again used for the wake to develop before averaging begins. 

\subsubsection{TNBL initialization}
\label{sssec:TNBL_init}

In the TNBL, also known as the Ekman layer \citep{ekman_influence_1905}, thermal stratification is absent throughout the entire domain. 
The TNBL potential temperature profile is initialized with $\theta(z) = \theta_0$, where $\theta_0 = \SI{300}{\kelvin}$ for all simulations. 
The boundary layer height is limited by the Rossby-Montgomery equilibrium height \citep{rossby_layer_1935, zilitinkevich_further_2007}. 
The domain height $L_z$ is selected such that the TNBL development is unaffected by the vertical domain constraint. 
We choose $L_z$ such that $L_z > 2.5h_E$ to minimize the domain dependence on TNBL development \citep{goit_effect_2015, jiang_large-eddy_2018}, where $h_E$ is the Rossby-Montgomery equilibrium height \citep{rossby_layer_1935}. 
For the $Ro = 500$ and $Ro = 1000$ simulations of the TNBL, the vertical domain is expanded to $L_z = 19.2D$ ($= \SI{4.6}{\kilo\meter}$) and $L_z = 38.4D$ ($=\SI{9.2}{\kilo\meter}$), respectively, to accommodate the boundary layer growth. 
The number of grid points is also increased to 384 and 768, respectively, to keep the vertical grid spacing $\Delta z$ equal to all other simulations.

\subsubsection{CNBL initialization}
\label{sssec:CNBL_init}

For simulations of the CNBL \citep{zilitinkevich_further_2007}, the ABL profile is initialized following the procedure of \citet{liu_geostrophic_2021}. 
An initial linear potential temperature profile $\theta(z) = \theta_0 + \Gamma z$ is used, where $\Gamma$ is the free atmosphere lapse rate. 
For all CNBL simulations, $\theta_0 = \SI{300}{\kelvin}$ and $\Gamma = \SI{1}{\kelvin\per\kilo\meter}$, simulating a weakly stratified free atmosphere \citep{sorbjan_effects_1996}. 
Unlike the TNBL simulations, the stable free atmosphere in the CNBL suppresses the ABL height and forms a weak capping inversion. 
A relatively weak lapse rate is chosen to mitigate the interaction of the turbine directly with the free atmosphere for low Rossby numbers, which will be discussed in \cref{ssec:results_waves}. 
Additionally, for the $Ro = 1000$ simulation in the CNBL, the vertical domain height is expanded to $L_z = 19.2D$ ($=\SI{4.6}{\kilo\meter}$) to accommodate boundary layer growth.

\section{Momentum budget analysis}
\label{sec:method_budgets}

To understand the structure and evolution of wakes in the ABL, we analyze the time-averaged streamwise and lateral momentum budgets. 
This will help to connect observations of wake recovery and wake deflection with the ABL forcings present in the governing equations. 
In \cref{ssec:RANS}, the time-averaged RANS equations are presented, and nomenclature is defined to refer to different forcing terms in the wake and ABL. 
Following this, in \cref{ssec:streamtubes}, a streamtube is defined to follow the wind turbine wake region, and the streamtube-averaged momentum budgets are presented to interpret the dynamics governing wake evolution and recovery. 

\subsection{Steady RANS equations}
\label{ssec:RANS}

To arrive at the filtered RANS equations, we apply the Reynolds decomposition to \cref{eq:NSE_LES} to separate the time-averaged $\overline{(\cdot)}$ and fluctuating components (\eg $u_i = \bar{u}_i + u_i'$). 
After applying the Reynolds decomposition, time-averaging yields 
\begin{equation}
    \label{eq:RANS}
    \bar{u}_j\pdv{\bar{u}_i}{x_j} 
    = 
    - \pdv{\bar{p}}{x_i} 
    + \frac{\delta_{i3}}{\theta_0 Fr^2} (\bar{\theta} - \theta_0) 
    - \pdv{\bar{\tau}_{ij}}{x_j}
    - \frac{1}{Ro}\varepsilon_{ij3}(G_j - \bar{u}_j)
    + \bar{f}_{t, i}
    - \pdv{\overline{u_i' u_j'}}{x_j}. 
\end{equation}
The final term in \cref{eq:RANS} is the divergence of the grid-resolved Reynolds stresses $\overline{u_i' u_j'}$, which can be interpreted as a turbulent flux of momentum. 

The wake recovery, quantified by the evolution of the streamwise velocity downwind of the turbine, is governed by the RANS equation in the streamwise ($i=1$) direction. 
In the wake away from the turbine, the turbine forcing $\bar{f}_{t, i}$ is zero and the streamwise RANS equations yield 
\begin{equation}
    \newcommand*{\vp}{\vphantom{\pdv{\overline{u_i'}}{x_j}}}
    \label{eq:U_RANS}
    \underbrace{ \vp
    \bar{u}_j \pdv{\bar{u}}{x_j}
    }_\text{I}
    = 
    \underbrace{ \vp
    -\pdv{\bar{p}}{x}
    }_\text{II}
    \underbrace{ \vp
    - \frac{1}{Ro} (G_2 - \bar{v})
    }_\text{III}
    \underbrace{ \vp
    - \pdv{\overline{u' u_j'}}{x_j}
    }_\text{IV}
    \underbrace{ \vp
    - \pdv{\bar{\tau}_{1j}}{x_j}
    }_\text{V}, 
\end{equation}
where $(u_1, u_2, u_3) = (u, v, w)$ is used to refer to the velocities in the streamwise, lateral, and vertical directions, respectively. 
The bracketed terms are I. Mean advection of streamwise momentum, II. Streamwise pressure gradient forcing, III. Direct Coriolis forcing, IV. Turbulent flux of streamwise momentum, and V. Divergence of subgrid stresses. 
The direct Coriolis forcing is defined in this study as the difference between the geostrophic pressure gradient and the Coriolis forcing term. 
The lateral geostrophic wind component $G_2 = G\sin(\alpha)$ is non-zero because the domain is rotated by angle $\alpha$ to align the hub-height inflow with the $x$ direction due to the Ekman spiral \citep{ekman_influence_1905}. 

Wake deflection is governed by the wake and ABL dynamics in the lateral ($i=2$) RANS equation. 
Lateral forcing in the wake leads to lateral wake acceleration, which induces nonzero $\bar{v}$ within the wake and causes wake deflection. 
The RANS equation for the lateral direction is given by 
\begin{equation}
\newcommand*{\vp}{\vphantom{\pdv{\overline{u_i'}}{x_j}}}
    \label{eq:V_RANS}
    \underbrace{\vp
    \bar{u}_j \pdv{\bar{v}}{x_j}
    }_\text{I} 
    = 
    \underbrace{\vp
    - \pdv{\bar{p}}{y}
    }_\text{II}
    \underbrace{\vp
    + \frac{1}{Ro}(G_1 - \bar{u})
    }_\text{III}
    \underbrace{\vp
    - \pdv{\overline{v' u_j'}}{x_j}
    }_\text{IV} 
    \underbrace{ \vp
    - \pdv{\tau_{2j}}{x_j}
    }_\text{V}
\end{equation}
where the bracketed terms are I. Mean advection of lateral momentum, II. Lateral pressure gradient forcing, III. Direct Coriolis forcing, IV. Turbulent flux of lateral momentum, and V. Subgrid stress divergence. 
In the canonical case of a turbulent wake in a pressure-driven boundary layer without Coriolis forcing, the flow is symmetrical across the lateral ($y$) direction and the lateral pressure gradient averages to zero across the wake region \citep{pope_turbulent_2000}. 
The addition of Coriolis forces breaks the wake symmetry and the lateral pressure gradients cannot necessarily be neglected. 

\subsection{Streamtube-averaged quantities}
\label{ssec:streamtubes}

\begin{figure}
    \centering
    \includegraphics[width=1.\linewidth]{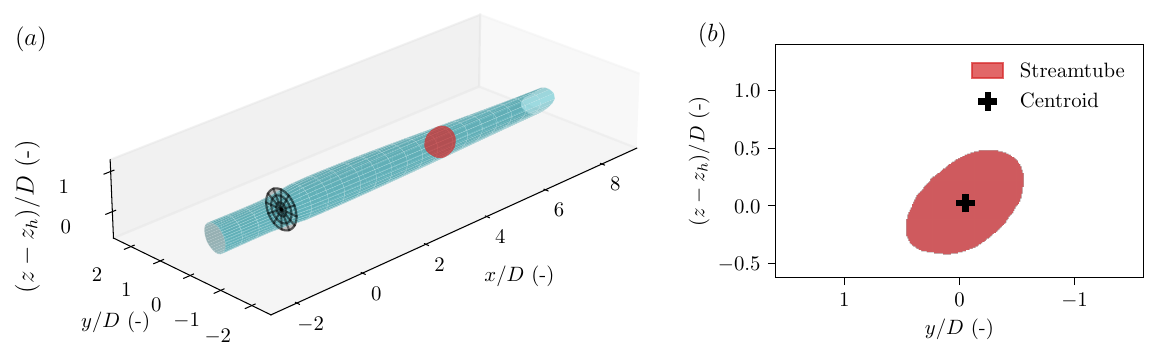}
    \caption{Schematic describing the streamtube boundary. ($a$) A streamtube for TNBL inflow at $Ro = 250$ wake is shown, seeded at the actuator disk with radius $r_s = 0.4D$. ($b$) A slice out of the streamtube shows the averaging region and centroid location. }
    \label{fig:streamtubes}
\end{figure}

To translate between observations of wake evolution and the wake momentum budgets, we define and track a streamtube containing the mass that intersects the rotor disk. 
The streamtube boundary is defined as the bundle of streamlines that passes through a ring of radius $r_s = 0.4D$ centered around and in the plane of the ADM \cite[\eg][]{shapiro_modelling_2018}. 
A schematic of a streamtube, the averaging region in the $yz$-plane, and the streamtube centroid position is shown in \cref{fig:streamtubes}. 
A larger streamtube ($r_s \sim 0.5D$) encapsulates a greater portion of the wake. 
In contrast, the wake dynamics and the background ABL flow, which is affected by wind shear, will be more homogeneous inside a smaller streamtube. 
Further, the edges of the streamtube for $r_s = 0.5D$ will be sensitive to the ADM regularization method, which smooths out the actuator disk forcing on the numerical grid \cite[]{shapiro_filtered_2019}.
As a compromise, we select $r_s = 0.4D$, but we note that the qualitative results are insensitive to the choice of initial streamtube radius, which is discussed in \cref{appx:streamtube_sensitivity}. 
Additionally, we present the wake centroid position $y_c(x)$ by computing the centroid of the streamtube cross-section, as shown in \cref{fig:streamtubes}($b$), but other definitions of the wake centroid \citep[c.f.\,][]{howland_wake_2016} yield the same qualitative results, which is also discussed in \cref{appx:streamtube_sensitivity}. 
Streamlines in the streamtube are computed by integrating the locus of starting points along the velocity field $\bar{u}_i(x, y, z)$. 
Because streamlines follow the the time-averaged flow, no mass crosses the streamtube boundary. 
As a result, the mass flux through any $yz$ cross-section of the streamtube (the region shown in \cref{fig:streamtubes}($b$)) is constant. 

% Volume-averaged budgets? 
In this work, we choose to study the wake dynamics by analyzing a streamtube seeded at the turbine rotor which follows the turbine wake. 
This choice is helpful for several reasons. 
First, the streamtube boundary represents a physically meaningful region that robustly follows the wake structure and evolution. 
The streamtube is an analog for the wake, tracking the wake position and wake deformation. 
Second, the momentum budgets can be connected to the observed wake behavior (\eg wake deflection) with the streamtube. 
For a given parcel of air in a Lagrangian frame of reference, a streamline trajectory can be recovered from an initial velocity and position by integrating the right-hand side of \cref{eq:RANS} in time, which is derived in \cref{appx:integration}. 
In this way, the trajectory of the parcel is an integrated form of the forcing terms in the momentum equations. 
The same principle applies to a bundle of streamlines which describes a streamtube. 
Third, the streamwise spatial evolution of field variables and budget quantities is easier to interpret for averaged quantities than for disaggregated profiles or planes. 
We choose to average quantities in the $yz$-plane within the streamtube to describe the mean dynamics within the wake as a function of the streamwise direction $x$, and $yz$-averaging within the streamtube is denoted $\langle \cdot \rangle$. 
Additionally, quantities that are averaged or integrated over the streamtube region are less sensitive to parameters in the streamtube definition (\eg where the streamtube is seeded) than over a rectangular domain enclosing the wake region, for which we find averaged quantities to be sensitive to both the choice of lateral and vertical domain boundaries. 
% This is because the streamtube evolves with the wake. 
For these reasons, streamtube-averaged quantities will be used to describe the evolution of the wake in the presence of Coriolis effects. 
Further discussion of the sensitivity of the reported wake dynamics results to various methods of post processing are given in  \cref{appx:streamtube_sensitivity}.

\section{Results}
\label{sec:results}

The results section begins in \cref{ssec:results_precursor} with the inflow profiles for uniform, TNBL, and CNBL conditions as a function of the relative Coriolis forcing strength, as quantified by the Rossby number. 
Then, we highlight the qualitative dependence of wake evolution for varying relative Coriolis strength in \cref{ssec:results_wakes}. 
The wake dynamics of the uniform inflow problem setup are analyzed in \cref{ssec:results_uniform}, before transitioning to the analysis of wake dynamics in ABL inflow in \cref{ssec:results_streamwise} and \cref{ssec:results_lateral}. 
Observations of internal gravity waves in low Rossby number CNBLs are discussed in \cref{ssec:results_waves}. 
Finally, we conclude the results section in \cref{ssec:results_implications} with a discussion of the statistical significance and relevance of Coriolis effects on wind turbine wakes in the context of wind farm design and control. 

\subsection{Precursor simulations}
\label{ssec:results_precursor}

\begin{figure}
    \centering
    \includegraphics[width=\linewidth]{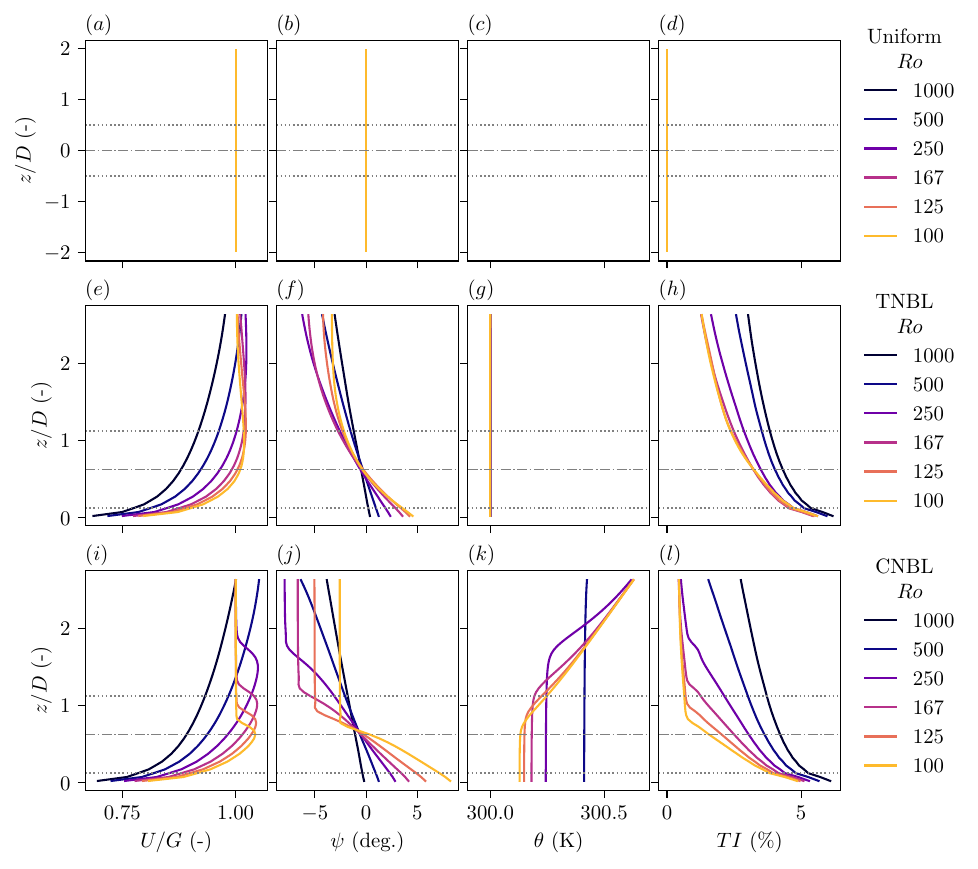}
    \caption{Horizontally and time-averaged inflow profile characteristics under ($a$-$d$) uniform, ($e$-$h$) truly neutral (TNBL), ($i$-$l$) and conventionally neutral (CNBL) conditions. 
    The turbine hub height $z_h$ is shown by the dash-dotted line, and the top and bottom extents of the rotor are shown by the black dotted lines. 
    Note that a prognostic equation for the potential temperature is not solved for the uniform inflow conditions.}
    \label{fig:inflow}
\end{figure}

Profiles for the horizontally averaged wind speed, wind direction, temperature, and turbulence intensity from the precursor simulations are shown in \cref{fig:inflow}. 
Turbulence intensity is defined as $TI = ((2/3) \langle \bar{e}^B \rangle_{xy})^{1/2} / U$, where $\bar{e}$ is the time-averaged, resolved turbulence kinetic energy, $\langle \cdot \rangle_{xy}$ denotes horizontal averaging, and the superscript $B$ denotes the base flow without turbines. 
The horizontally-averaged wind speed magnitude is given by $U = (\langle \bar{u}^B \rangle_{xy}^2 + \langle \bar{v}^B\rangle_{xy}^2)^{1/2}$. 
In the uniform inflow simulations, shown in \cref{fig:inflow}($a$-$d$), inflow profiles are independent of the Rossby number. 
This is due to the absence of velocity shear in uniform inflow and a domain with slip-walls. 
Note that the vertical domain extent is much larger than the profiles shown in \cref{fig:inflow}, which are zoomed in to the rotor area. 

Adding a bottom wall introduces velocity shear and thus vertical variation in the inflow profiles as a function of the Rossby number for the ABL simulations, shown in \cref{fig:inflow}($e$-$l$). 
In the TNBL, the wind speed magnitude $U$ increases to a maximum of $U_{\max} \approx 1.02G$ at the low-level jet. 
The non-dimensional friction velocity $u_*/G$ shows a weak inverse dependence on the Rossby number, and a summary of inflow and ABL properties is given in \cref{tab:inflow_properties}. 
The height of the ABL depends on the Rossby number by the Rossby-Montgomery equilibrium height $h_E \sim u_*/|f_c|$ \citep{rossby_layer_1935, zilitinkevich_further_2007}. 
In \cref{tab:inflow_properties}, we compute the ABL height $h$ as the vertical height where the turbulent fluxes reach 5\% of their surface value \citep{zilitinkevich_further_2007}. 
The magnitude of wind direction shear at the surface monotonically increases with increasing relative Coriolis strength (decreasing $Ro$). 
However, wind direction profiles are increasingly non-linear in the rotor area as the TNBL height decreases with decreasing Rossby number. 
The change in time-mean wind direction from the bottom to the top of the rotor is defined as $\Delta \winddir = \winddir(z_h+D/2) - \winddir(z_h-D/2)$ for the time-averaged wind direction $\winddir(z) = \arctan(\langle \bar{v}^B \rangle_{xy}/ \langle \bar{u}^B \rangle_{xy})$. 
Overall, due to the presence of the bottom boundary, the TNBL inflow is significantly more complex than the uniform inflow simulations. 

In the CNBL, the boundary layer growth is suppressed relative to the TNBL due to the presence of the stable free atmosphere. 
As in the TNBL, the jet height and ABL height descend with decreasing $Ro$ following the geostrophic drag law \citep{zilitinkevich_further_2007, liu_geostrophic_2021}. 
The low-level jet is stronger in the CNBL than in the TNBL and the maximum wind speed reaches $1.05G$. 
The non-dimensional friction velocity again shows a weak inverse dependence on the relative strength of Coriolis forcing, as shown in \cref{tab:inflow_properties}, which is consistent with previous simulations of the CNBL \citep{liu_geostrophic_2021}. 
The CNBL heights are smaller than the TNBL height for the same Rossby numbers. 
Additionally, in the low Rossby number simulations ($Ro \leq 167$), the CNBL is sufficiently shallow that the turbine, which reaches up to \SI{270}{\meter} at the rotor tip, impinges on the free atmosphere. 
The effects of the rotor interacting directly with the free atmosphere, which is a consequence of the Rossby number, free atmosphere lapse rate, and surface roughness, are discussed in \cref{ssec:results_waves}. 
Again, we note that a relatively weak free atmosphere lapse rate is used in all CNBL simulations to promote ABL growth. 
Similar to the TNBL, there is a non-monotonic trend in the wind direction profiles across the turbine rotor $\Delta \winddir$. 
That is, the change in wind direction across the rotor does not monotonically increase with increasing relative Coriolis forcing strength. 
Wind direction shear across the rotor of the turbine $\Delta \winddir$ spans a larger range of values for the same parametric sweep of $Ro$ in the CNBL than in the TNBL (see \cref{tab:inflow_properties}). 

\begin{table}
\centering
\caption{Inflow properties of the LES experiments in this study. }
\footnotesize
\begin{tabular}{lllllllll}
\hline
Flow type & $G$ (m s$^{-1}$) & $Ro$ (-) & $u_*/G$ (-) & $h$ (m) & $\winddir_{hub}~(^\circ)$ & $u_{hub}/G$ (-) & $U_{max}/G$ (-) & $\Delta \winddir~(^\circ)$ \\ \hline
\multirow{6}{*}{Uniform}
 & 24.7 & 1000 & - & - & 0 & 1 & 1 & 0 \\
 & 12.4 & 500 & - & - & 0 & 1 & 1 & 0 \\
 & 6.2 & 250 & - & - & 0 & 1 & 1 & 0 \\
 & 4.1 & 167 & - & - & 0 & 1 & 1 & 0 \\
 & 3.1 & 125 & - & - & 0 & 1 & 1 & 0 \\
 & 2.5 & 100 & - & - & 0 & 1 & 1 & 0 \\ \hline
\multirow{5}{*}{TNBL} 
 & 24.7 & 1000 & 0.0252 & 3057 & -0.50 & 0.88 & 1.02 & -1.4 \\
 & 12.4 & 500 & 0.0264 & 1673 & -0.32 & 0.92 & 1.02 & -2.5 \\
 & 6.2 & 250 & 0.0274 & 831 & -0.56 & 0.97 & 1.02 & -4.5 \\
 & 4.1 & 167 & 0.0283 & 604 & -0.41 & 0.99 & 1.02 & -5.6 \\
 & 3.1 & 125 & 0.0288 & 493 & -0.28 & 1.00 & 1.02 & -5.7 \\
 & 2.5 & 100 & 0.0293 & 477 & -0.37 & 1.01 & 1.02 & -5.7 \\ \hline
\multirow{5}{*}{CNBL} 
 & 24.7 & 1000 & 0.0250 & 1359 & -1.07 & 0.89 & 1.05 & -1.5 \\
 & 12.4 & 500 & 0.0263 & 717 & -0.60 & 0.94 & 1.05 & -3.0 \\
 & 6.2 & 250 & 0.0277 & 390 & -0.48 & 0.98 & 1.05 & -5.4 \\
 & 4.1 & 167 & 0.0285 & 272 & -0.50 & 1.01 & 1.05 & -8.9 \\
 & 3.1 & 125 & 0.0291 & 210 & -0.23 & 1.03 & 1.05 & -9.9 \\
 & 2.5 & 100 & 0.0296 & 174 & 0.35 & 1.04 & 1.04 & -9.7 \\ \hline
\end{tabular}
\label{tab:inflow_properties}
\end{table}

\subsection{Qualitative time-averaged wake behavior}
\label{ssec:results_wakes}

\begin{figure}
    \centering
    \includegraphics[width=\linewidth]{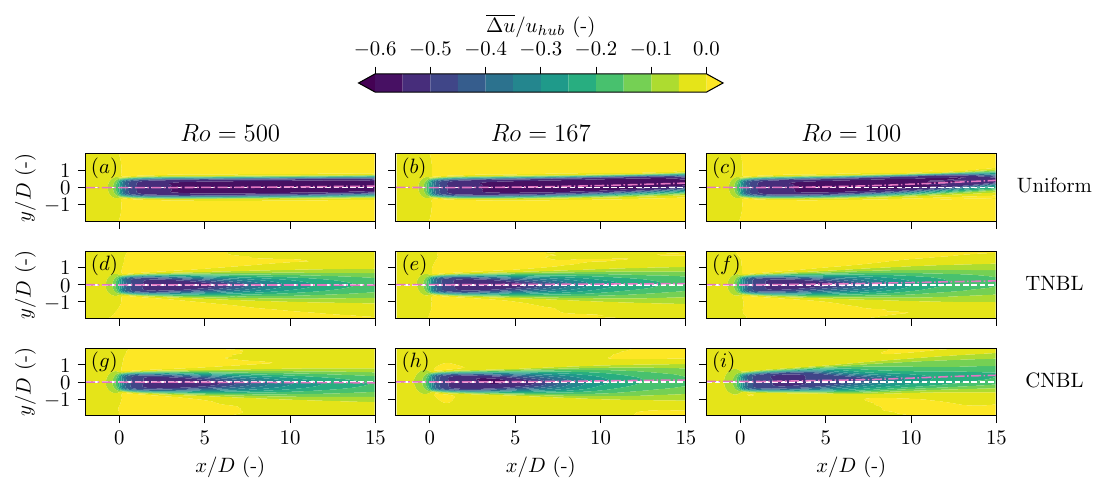}
    \caption{Hub height wind turbine wakes visualized as the streamwise velocity deficit with respect to the inflow, viewed from above, for varying Rossby numbers in ($a$, $b$, $c$) uniform inflow, ($d$, $e$, $f$) TNBL inflow, and ($g$, $h$, $i$) CNBL inflow. The dashed white line indicates $y=0$ while pink dash-dotted lines show the wake centerline. }
    \label{fig:xy_wakes}
\end{figure}

In this section, we examine the time-averaged wake velocity fields. 
Cross sections of the hub height velocity deficit field $\overline{\Delta u}_i = \bar{u}_i - \langle \bar{u}_i^B \rangle_{xy}$ in the streamwise direction ($i=1$) are shown in \cref{fig:xy_wakes}, where $\langle \bar{u}_i^B \rangle_{xy}$ is the horizontally-averaged, time-averaged precursor flow. 
The precursor flow is the base flow without wakes.
In uniform inflow, wake recovery does not begin until very far downstream ($x/D \gtrsim 15$) due to the lack of freestream turbulence \citep{howland_wake_2016}. 
Wake deflection is anti-clockwise as viewed from above (deflection is in the $+y$ direction in the chosen coordinate system). 
The amount of wake deflection increases with increasing relative Coriolis forcing strength (\ie decreasing $Ro$). 

For the TNBL and CNBL inflows, wakes recover due to freestream turbulence. 
The wake recovery in all simulations is qualitatively only minimally affected by the parametric effects of $Ro$, despite the large variation in turbulence intensity, which is dependent on $Ro$, across the rotor plane. 
Wake deflection is also increasingly anti-clockwise as the relative strength of Coriolis forcing increases, which is particularly noticeable in the CNBL simulations at low Rossby numbers. 
A quantitative investigation of the wake dynamics in the ABL flows is reserved for \cref{ssec:results_streamwise} and \cref{ssec:results_lateral}. 

\begin{figure}
    \centering
    \includegraphics[width=0.8\linewidth]{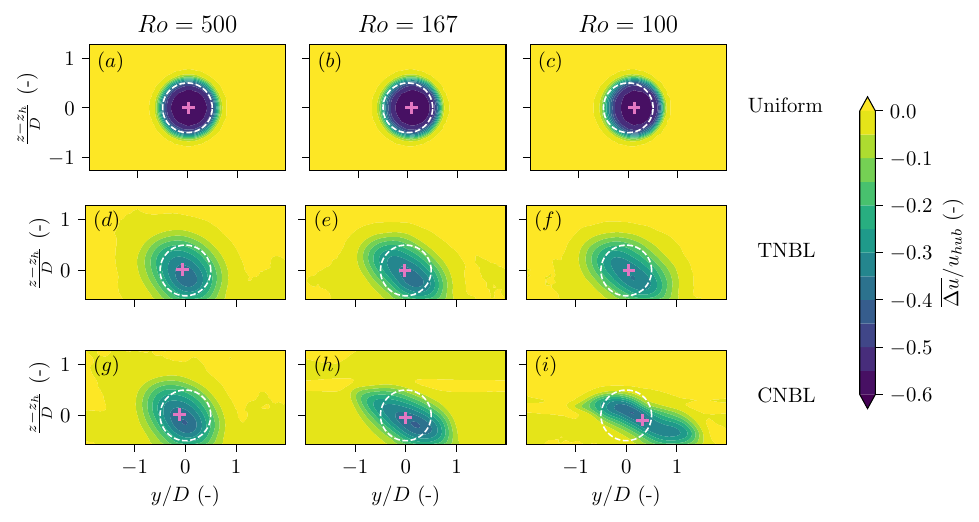}
    \caption{Wake cross-sections at $x/D = 8$ visualized as the streamwise velocity deficit with respect to the inflow for varying Rossby numbers in ($a$, $b$, $c$) uniform inflow, ($d$, $e$, $f$) TNBL inflow, and ($g$, $h$, $i$) CNBL inflow. The turbine location is given by the white circle centered around the origin, and the wake centroid position is given by the pink $+$. }
    \label{fig:yz_wakes}
\end{figure}

Vertical slices through the $yz$-plane are shown in \cref{fig:yz_wakes} at a distance $x = 8D$ downwind of the rotor.  
By $x=8D$, variations in the relative Coriolis forcing strength can be observed in the uniform inflow wakes, with lower Rossby numbers deflecting farther in the $+y$ direction.
In the TNBL, wakes are skewed, which is a result of the inflow wind direction shear \citep{magnusson_influence_1994, abkar_influence_2016}. 
The difference in wake skewing between Rossby numbers is small, owing to the small spread in wind direction change over the rotor $\Delta \winddir$ for the TNBL simulations. 

In the CNBL, the decreasing boundary layer height as the Rossby number decreases strongly affects the wake shape. 
The large differences in wake shape between the CNBL simulations result from the changing inflow properties. 
Because the inflow wind direction is increasingly non-linear with decreasing $Ro$, wake velocity slices assume complex shapes that are not well approximated by axisymmetric or elliptical (skewed) Gaussian wakes \citep{abkar_influence_2016}. 
Additionally, the wind speed shear across the rotor is increasingly complex as the ABL height decreases and the rotor interacts directly with the low-level jet. 
Finally, the ambient turbulence intensity varies substantially across the rotor, particularly for $Ro = 125$ and $Ro = 100$, which is distinct from the TNBL simulations. 
% As a result, the CNBL wakes do not follow an axisymmetric Gaussian or even a skewed Gaussian wake shape. 

\subsection{Uniform inflow results}
\label{ssec:results_uniform}

The uniform inflow simulations represent the canonical flow building-block for understanding the dynamics of turbine wakes in the presence of Coriolis forces. 
Specifically, the uniform inflow environment separates Coriolis forcing from wind shear. 
Although the presence of uniform (\ie gradient-free and non-turbulent) inflow cannot exist in the ABL due to the presence of the ground, it is helpful to study as a basis for parsing the dominant dynamics governing wake structure and evolution in more complex inflows. 
The main limitation of the uniform inflow numerical setup is the unphysical delay in the onset of a fully-turbulent far wake, as shown in \cref{fig:xy_wakes}. 
Therefore, we will not attempt to draw conclusions from the streamwise momentum budgets regarding wake recovery using the uniform inflow setup. 
Nonetheless, we will show that the dynamics of the uniform inflow problem are complex and instructive for parsing mechanisms of momentum transport affects wake evolution in the presence of Coriolis forces. 

\begin{figure}
    \centering
    \includegraphics[width=\linewidth]{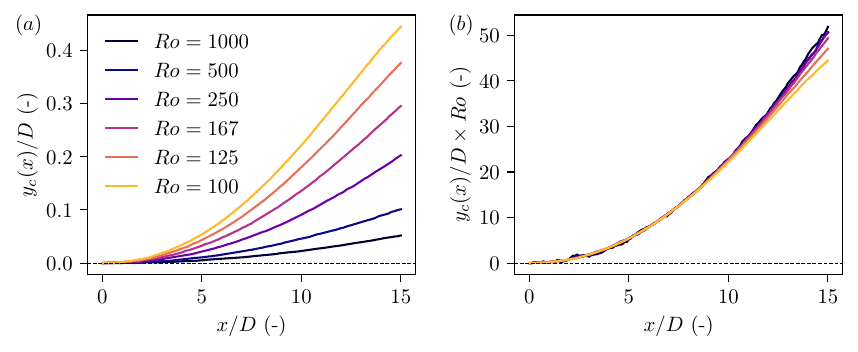}
    \caption{($a$) Streamtube centroid position $y_c(x)$ as a function of downstream distance $x$ in uniform inflow varying the Rossby number. ($b$) Scaling by the Rossby number collapses all uniform inflow simulations onto one curve. }
    \label{fig:coriolis_similarity}
\end{figure}

We begin by investigating the wake deflection due to the Coriolis force in uniform inflow. 
Here, we choose to measure the wake deflection $y_c(x)$ as the centroid of the streamtube cross-section in the $yz$-plane. 
As discussed in \cref{ssec:streamtubes}, the wake deflection based on alternative definitions of the wake centroid show the same qualitative trends, shown in \cref{appx:streamtube_sensitivity}. 
The wake deflection increases monotonically with increasing relative Coriolis forcing strength (decreasing $Ro$) and follows a parabolic trajectory, as shown in \cref{fig:coriolis_similarity}($a$). 
By scaling the wake deflection $y_c$ by the Rossby number, shown in \cref{fig:coriolis_similarity}($b$), we see that the wake deflection is self-similar until $x/D\approx 12$. 
The departure from self-similarity for $x/D > 12$ is associated with wake breakdown and turbulence.
The self-similarity allows us to focus on understanding the wake dynamics at one Rossby number with the expectation that the relevant physical transport mechanisms will scale with the Rossby number. 

\begin{figure}
    \centering
    \includegraphics[width=0.8\linewidth]{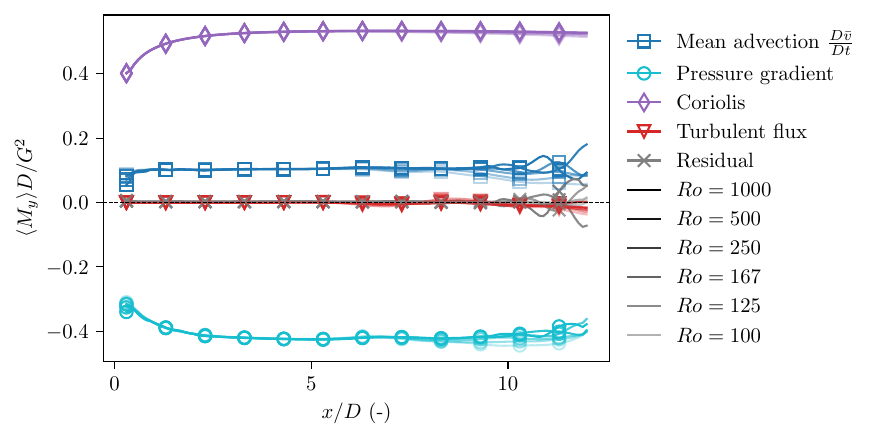}
    \caption{All terms in the lateral momentum balance for the uniform inflow simulations, streamtube-averaged and plotted as a function of streamwise distance. The streamtube-averaged quantities are scaled by the Rossby number $Ro$, showing self-similarity in the flow. }
    \label{fig:uniform_ymomentum}
\end{figure}

The wake deflection is the integrated form of the lateral momentum budget. 
Therefore, in uniform inflow, the lateral momentum budgets should also be self-similar if scaled by the Rossby number. 
Individual terms $M_y$ in the streamtube-averaged lateral momentum balance \cref{eq:V_RANS} are shown in \cref{fig:uniform_ymomentum} as a function of streamwise position downwind of the turbine. 
The individual terms, when scaled by the Rossby number, are also self-similar. 
All of the forcing terms are nearly constant in $x$ outside of the wake expansion region $x/D \lesssim 1$ until wake recovery begins at $x/D\approx 12$. 
Therefore, the uniform inflow centroid position follows a parabolic trajectory due to the constant acceleration (mean advection). 
The Coriolis forcing term is the largest in magnitude and positive in sign (in the northern hemisphere), which results in a positive lateral acceleration and therefore a net positive lateral advection of the wake region. 
However, the presence of a lateral pressure gradient opposes the direct Coriolis forcing term. 
The result of the pressure gradient is a net acceleration which is only about 25\% as strong as the direct Coriolis forcing term. 

\begin{figure}
    \centering
    \includegraphics[width=\linewidth]{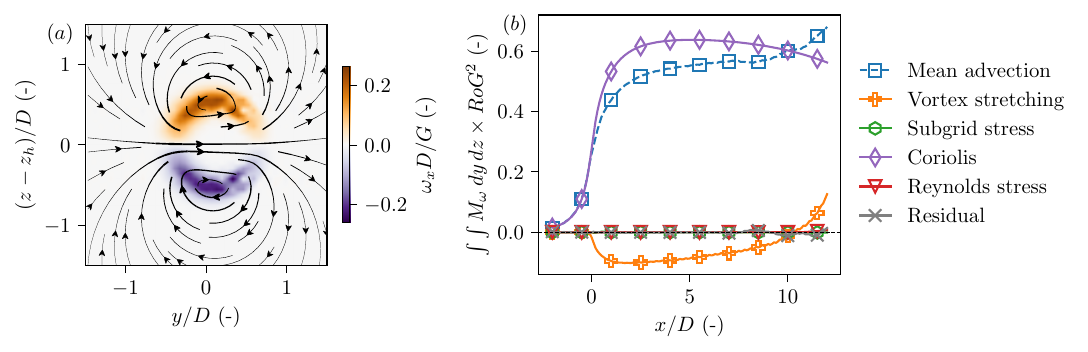}
    \caption{($a$) Streamlines of in-plane velocities $\bar{v}$ and $\bar{w}$ induced by the Coriolis force. Contours of vorticity are superimposed, showing a CVP with positive streamwise vorticity above the hub plane and negative vorticity below. ($b$) Streamwise vorticity budget terms $M_\omega$, integrated for $z > z_h$. Streamwise vorticity advection (dashed blue) is balanced by Coriolis production (purple).}
    \label{fig:coriolis_vorticity}
\end{figure}

% It is not trivial why there is a strong lateral pressure gradient which persists far downstream. 
We seek to investigate the physical mechanism that results in a non-zero lateral pressure gradient. 
The streamtube averaged lateral pressure gradient in the absence of Coriolis forcing is equal to zero.
To investigate how Coriolis effects give rise to the lateral pressure gradients, we visualize streamlines of the in-plane velocity components $\bar{v}, \bar{w}$ at a cross-section $x/D = 10$. 
In \cref{fig:coriolis_vorticity}($a$), streamlines of the in-plane velocity components show the presence of a counter-rotating vortex pair (CVP), which is the source of the wake deflection. 
The formation of the CVP, which is visualized by the formation of streamwise vorticity $\omega_1 = \omega_x$, can be understood through the vorticity budget equation. 
We take the curl of \cref{eq:RANS} to solve for the Reynolds-averaged vorticity equation. 
% in the streamwise ($i=1$) direction.  
This yields
\begin{equation}
\label{eq:RANS_vort}
\newcommand*{\vp}{\vphantom{\left(\pdv{\overline{u'_k}}{\bar{A}_j}\right)}}
    \underbrace{\vp 
    \bar{u}_j \pdv{\bar{\omega}_i}{x_j}
    }_\text{A}
    = 
    \underbrace{\vp 
    \bar{\omega}_j \pdv{\bar{u}_i}{x_j}
    }_\text{B}
    \underbrace{\vp 
    + \frac{\varepsilon_{ij3}}{Fr^2 \theta_0} \pdv{\bar{\theta}}{x_j}
    }_\text{C}
    \underbrace{\vp 
    - \varepsilon_{ijk} \pdv{}{x_j} \left( 
    \pdv {\bar{\tau}_{km}}{x_m} + \pdv {\overline{u_k' u_m'}}{x_m}
    \right)
    }_\text{D}
    \underbrace{\vp 
    + \frac{1}{Ro} \pdv{\bar{u}_i}{x_3}
    }_\text{E}, 
\end{equation}
where the bracketed terms are A. Mean advection due to mean flow, B. Vortex stretching, C. Buoyancy torque, D. Torque from subgrid and Reynolds stresses, and E. Coriolis. 
The streamwise vorticity causes the CVP formation in uniform inflow. 
Examining the streamwise vorticity budget ($i=1$), the dominant terms in \cref{fig:coriolis_vorticity}($b$) show that the transfer of vorticity from Coriolis forces is primarily balanced by the mean advection of vorticity. 
This balance can be written as 
\begin{equation}
\label{eq:wx_vorticity}
    \frac{D\bar{\omega}_x}{Dt} 
    \approx 
    \bar{u}\frac{\partial \bar{\omega}_x}{\partial x} 
    \approx
    \frac{1}{Ro} \frac{\partial \bar{u}}{\partial z}. 
\end{equation}
In the vorticity balance, the role of the Coriolis force is to introduce planetary vorticity into the wake dynamics. 
The presence of the wake induces velocity shear such that positive streamwise vorticity is generated above hub height (in the northern hemisphere), and vice-versa below hub height. 
% $\partial \bar{u}/\partial z$ is positive above hub height, leading to positive streamwise vorticity generation above hub height in the northern hemisphere. 

Lateral velocities induced by a CVP are also the mechanism for wake deflection in yaw-misaligned turbines \citep{howland_wake_2016, bastankhah_experimental_2016}. 
Wakes trailing yawed turbines are a widely studied case in recent literature \citep{meyers_wind_2022}, and therefore we make an analogy between the flow physics from Coriolis effects and yawed turbine wakes here. 
Traditionally, modeling the deflection of yaw-misaligned wind turbine wakes with momentum-based approaches has yielded overpredictions of the wake deflection because the lateral pressure gradient forces on the streamtube, which are non-negligible, are neglected for model simplicity \citep{jimenez_application_2010, shapiro_modelling_2018, heck_modelling_2023}. 
For wakes in uniform inflow with Coriolis forcing, the presence of a CVP also induces a non-negligible lateral pressure gradient force. 
The role of the pressure gradient forcing, as is the case in yaw-misaligned wind turbine wakes, is to enforce mass conservation \citep{bastankhah_experimental_2016}. 
Neglecting the pressure gradient forcing in \cref{eq:V_RANS} leads to an order-of-magnitude overprediction in the wake deflection (not shown). 

In summary, the net effect of the Coriolis force on wakes in uniform inflow is an anti-clockwise deflection (as viewed from above) that increases with increasing relative Coriolis forcing strength. 
In the lateral momentum budget, the Coriolis term is dominant but partially cancelled by a lateral pressure gradient force in the wake. 
Wake deflection is caused by the formation of a CVP due to the transfer of planetary vorticity into the streamwise direction via the wake-added shear. 
The CVP is responsible for creating $\bar{v} > 0$ inside the wake (in the northern hemisphere), which deflects the wake anti-clockwise and gives rise to the non-negligible lateral pressure gradient force. 
Neglecting the pressure gradient forcing in a momentum-based modeling approach results in an overprediction in the wake deflection. 

\subsection{Coriolis effects on streamwise momentum in the ABL}
\label{ssec:results_streamwise}

\begin{figure}
    \centering
    \includegraphics[width=\linewidth]{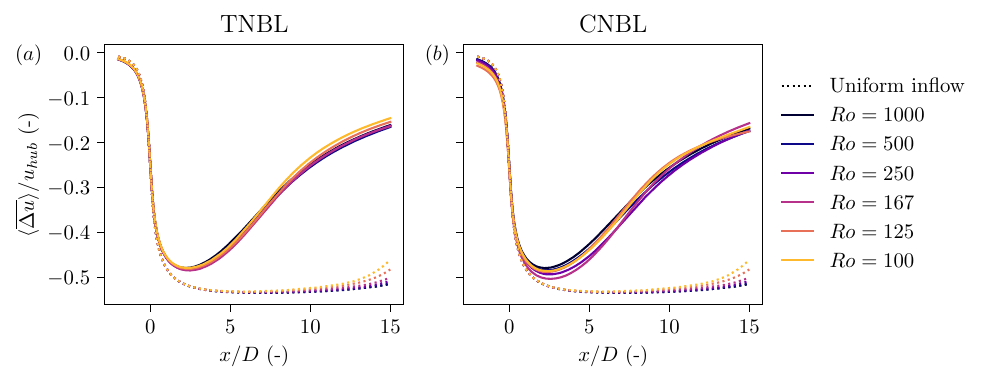}
    \caption{Streamtube-averaged streamwise velocity deficit as a function of streamwise coordinate $x/D$ in $(a)$ TNBL inflow and $(b)$ CNBL inflow. Uniform inflow wakes are shown with dotted lines. }
    \label{fig:wake_deficit}
\end{figure}

This section focuses on wake recovery in TNBL and CNBL conditions.
The streamtube-averaged streamwise velocity deficit is shown as a function of streamwise distance in \cref{fig:wake_deficit}. 
In the uniform inflow cases, shown as dotted lines in \cref{fig:wake_deficit}, the streamtube velocity deficit is constant after the pressure recovery in the near-wake of the turbine. 
Additionally, the uniform inflow wake recovery does not depend on the strength of Coriolis forcing. 
Overall, in uniform inflow, we observe very weak parametric dependence of relative Coriolis forcing strength on wake recovery. 

In the TNBL, the relative strength of Coriolis forcing weakly affects the wake recovery. 
Due to wind shear in the ABL inflow, the turbine thrust in the TNBL is lower than in uniform inflow. 
As a result, the wake velocity deficit magnitude is lower in the TNBL than in uniform inflow. 
In the far wake, the strongest relative Coriolis forcing experiences the fastest wake recovery, and the wake recovery rate generally decreases as $Ro$ increases. 
We note that the relative difference in the wake strength $\langle \overline{\Delta u}\rangle$ between the strongest and weakest Coriolis forcing is approximately 12\% at $x=15D$. 
We further elaborate on the statistical significance of the differences in wake recovery for varying $Ro$ in \cref{ssec:results_implications}. 

% In contrast, the CNBL shows complex behavior in the wake recover. 
Trends in wake velocity as a function of $Ro$ are less clear in the CNBL. 
This is due to the suppressed ABL height, which results in complex inflow conditions across the rotor area, as noted in \cref{ssec:results_wakes}. 
For example, the wind speed profile changes significantly as the boundary layer height decreases and the low-level jet interacts with the rotor. 
This affects the rotor thrust, which increases monotonically with decreasing CNBL height. 
As a result, the maximum velocity deficit, which occurs around $x \approx 2.5D$, changes in magnitude with the Rossby number. 
Further, the turbulence intensity also varies more in the CNBL than in the TNBL as a function of the Rossby number. 
In contrast with the TNBL wakes, wake recovery rates in the CNBL do not follow a clear trend with changing Rossby number. 
To parse the dynamics of the wake recovery, we analyze the streamwise momentum budget in the wake. 

\begin{figure}
    \centering
    \includegraphics[width=1.\linewidth]{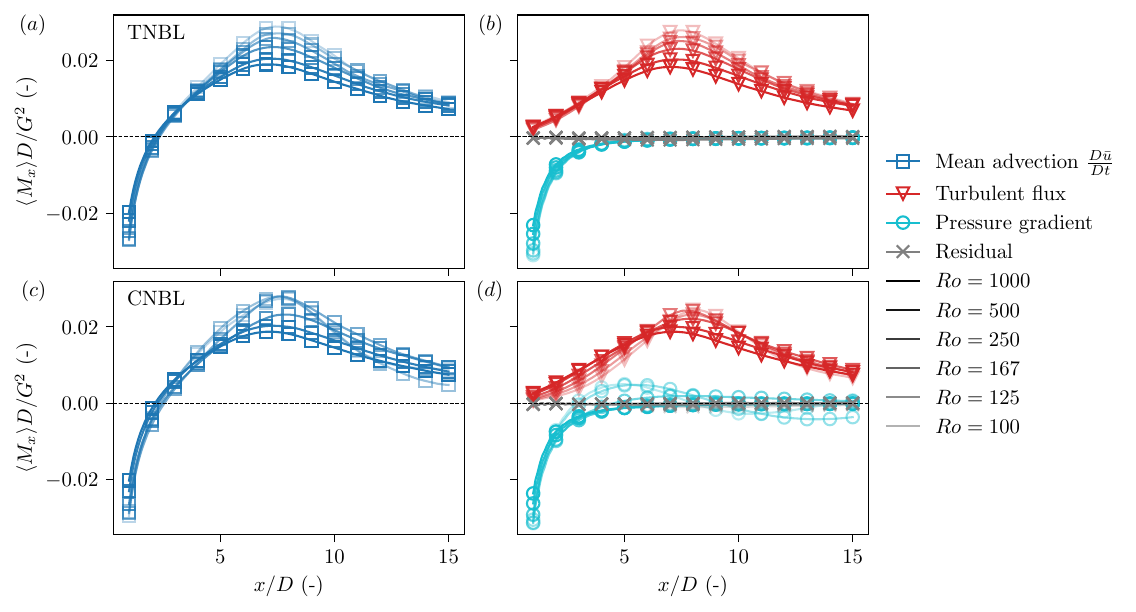}
    \caption{Streamtube-averaged streamwise momentum budget terms $M_x$ for wakes in the ($a$, $b$) TNBL and ($c$, $d$) CNBL. The mean advection is the sum of all forcing terms. }
    \label{fig:streamtube_xmom}
\end{figure}

Wake recovery is described by the streamwise momentum budget. 
The streamtube-averaged, streamwise momentum budget is shown in \cref{fig:streamtube_xmom} for wakes in the TNBL and CNBL. 
For $x \lesssim 3D$, the pressure gradient term is large due to the pressure recovery from the forcing at the rotor. 
Beyond this region, turbulent fluxes (divergence of Reynolds stresses) replenish momentum into the wake, comprising the dominant dynamics. 

In the TNBL, the wake recovery due to the turbulent flux of momentum accounts for nearly all (95-99\%) of the wake recovery. 
Effects of direct Coriolis forcing, subgrid stresses, and pressure gradients are negligible in the streamwise momentum budget. 
In other words, the dependence of the streamwise momentum budget on relative Coriolis forcing strength is primarily through changes in the ABL structure, which depends on the Rossby number (see \cref{fig:inflow}). 
The Reynolds stress divergence in the wake, and therefore wake recovery rate, increases as the relative strength of Coriolis forcing increases. 
For all Rossby numbers, the Reynolds stress divergence reaches a maximum value near $x = 8D$. 
The marginal increase in wake recovery rate with decreasing Rossby number shown in \cref{fig:wake_deficit}($a$) is enabled by the increasing turbulent flux in the wake. 

We observe two notable differences between the TNBL and CNBL streamwise momentum budgets. 
First, the inflow turbulence intensity at the rotor hub-height increases more with increasing $Ro$ in the CNBL than in the TNBL. 
As a result, the turbulence flux term in \cref{fig:wake_deficit}($b$) monotonically increases with increasing Rossby number for $x \lesssim 5D$. 
The peak Reynolds stress divergence still generally increases with increasing relative Coriolis forcing (decreasing $Ro$), as was observed in the TNBL. 
The exception is for the cases where stable thermal stratification in the free atmosphere suppresses turbulence and therefore wake recovery due to the shallow CNBL height ($Ro = 125$ and $Ro = 100$). 
% $Ro = 100$, which experiences the strongest suppression of wake recovery due to stable thermal stratification from the decreasing free atmosphere height. 
Second, the pressure gradient forcing deviates from the asymptotic recovery behavior expected from classical momentum theory. 
For $Ro \leq 167$, we instead observe oscillations in the pressure gradient term. 
These oscillations stem from gravity waves induced by the free-standing turbine, which are a product of ABL conditions at low Rossby numbers and are elaborated on in \cref{ssec:results_waves}. 
The gravity wave pressure oscillations cause a favorable pressure gradient in some regions of the turbine wake which temporarily increases the velocity recovery within the wake, as observed in \cref{fig:wake_deficit}($b$). 

\begin{figure}
    \centering
    \includegraphics[width=\linewidth]{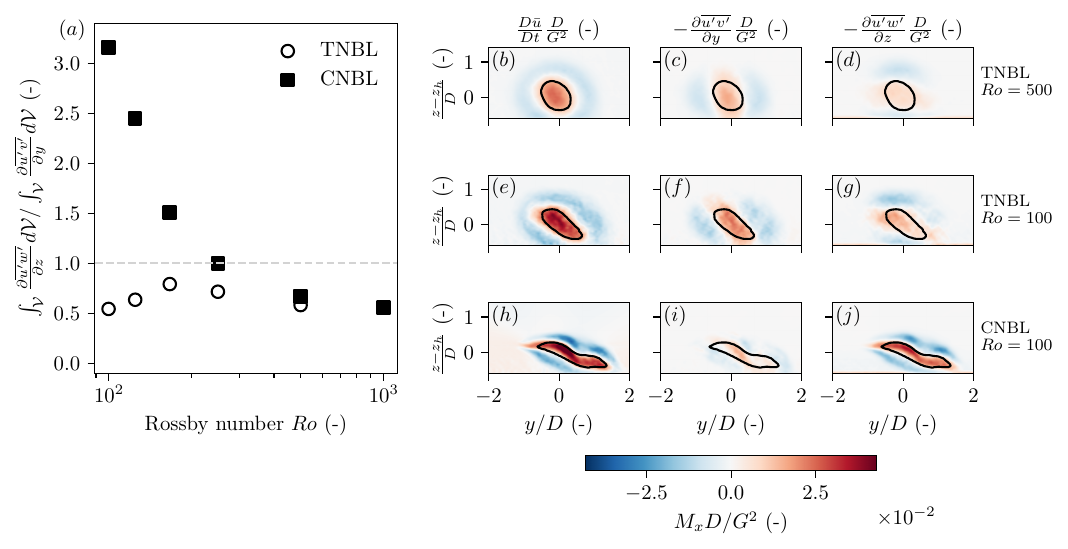}
    \caption{($a$) Ratio of vertical to lateral turbulent momentum flux into the wake streamtube. ($b$-$j$) Profiles in the $yz$-plane at $x = 8D$ of gradients of turbulent momentum flux, where reds indicate acceleration (wake recovery) and blues indicate deceleration. The streamtube boundary is outlined in black. }
    \label{fig:streamwise_recovery}
\end{figure}

To examine the parametric dependence of the turbulent momentum flux on the Rossby number, we take a further look at the Reynolds stress divergence term. 
In \cref{fig:streamwise_recovery}, we break out the wake recovery due to the Reynolds stress divergence into lateral ($M_{x,2}$) and vertical ($M_{x,3}$) turbulent momentum fluxes. 
The streamwise Reynolds stress component is not shown because it is small compared to the in-plane components and integrates to nearly zero inside the streamtube for $x \in [0, 15]D$. 

% Maybe this would be better divided into polar coordinates?

Both the magnitude and structure of turbulent fluxes change as a function of $Ro$. 
In \cref{fig:streamwise_recovery}($a$), the ratio of vertical to lateral turbulent momentum flux $M_{x,3}/M_{x,2}$, integrated over the entire streamtube from the rotor to $x = 15D$, are shown as a function of $Ro$. 
In the present simulations of a single turbine wake in the TNBL, the vertical momentum flux contribution is between 55-80\% of the lateral momentum flux contribution to the wake recovery. 
While the ratio of vertical to lateral momentum flux changes modestly with $Ro$ in the TNBL, the dependence of $M_{x,3}/M_{x,2}$ on $Ro$ in the CNBL is more pronounced. 
In the CNBL, as the relative Coriolis forcing strength increases and ABL height decreases, wake recovery is increasingly driven by vertical momentum fluxes. 
We note that while the ratio of turbulent fluxes is dependent on ABL properties and Rossby number, as shown here, it is also dependent on lateral turbine spacing if neighboring turbines are present \citep{calaf_large_2010, van_der_laan_brief_2023}. 
% Therefore, wake recovery within wind farm arrays may be more sensitive to changes in $Ro$ than the single turbine wake we study here. 
Therefore, the dependence of wake recovery on Rossby number may differ for turbine wakes in wind farms compared to the wakes of free-standing turbines studied here. 

We visualize turbulent momentum flux contributions to the wake recovery in \cref{fig:streamwise_recovery}($b$-$j$). 
Contours of the mean streamwise advection, lateral Reynolds stress gradient, and vertical Reynolds stress gradient are shown in a $yz$-plane at a distance $x=8D$ downwind for three atmospheric conditions. 
In all cases, the streamtube boundary corresponds closely to the boundary between net Reynolds stress convergence (acceleration) and divergence (deceleration). 
That is, the streamtube-averaged quantities robustly select the wake region which is accelerating and recovering lost momentum. 
At $Ro = 500$, the TNBL and CNBL wake recovery is very similar, as shown by \cref{fig:streamwise_recovery}($a$), so only the contours of the TNBL wake are shown. 
% In contrast, we find that the same ratio of turbulent fluxes is quite sensitive to choice of control volume extents if a rectangular control volume is used. 
Between $Ro = 500$ and $Ro = 100$ in the TNBL and CNBL, the magnitude of the mean advection within the streamtube increases. 
In the TNBL, changes in the Rossby number only weakly affect the wake shape and the ABL structure, so the individual turbulent flux contributions are similar between $Ro = 100$ and $Ro = 500$. 
In the CNBL, wakes experience strong wind direction shear as the ABL height decreases with decreasing $Ro$. 
This strongly skews the wake and causes more vertical momentum entrainment than lateral entrainment of streamwise momentum. 
To summarize, while differences in overall wake recovery are small between Rossby numbers, wake recovery mechanisms change substantially as the ABL height decreases.

\subsection{Coriolis effects on lateral momentum}
\label{ssec:results_lateral}

\begin{figure}
    \centering
    \includegraphics[width=\linewidth]{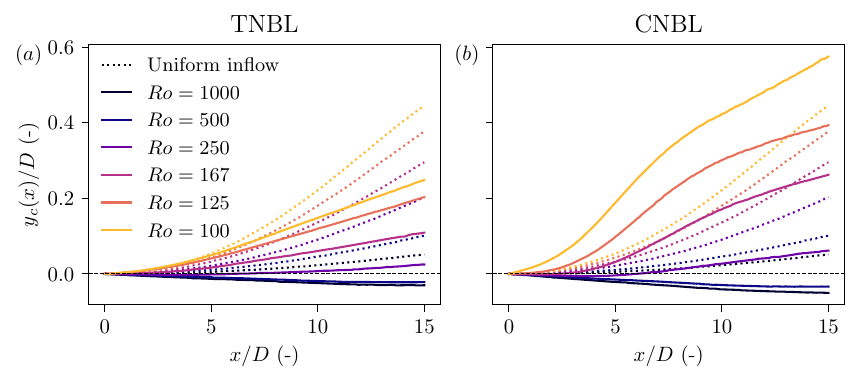}
    \caption{Wake centroid location, defined as the centroid of the streamtube, as a function of streamwise coordinate $x/D$ for varying Rossby numbers in the ($a$) TNBL and ($b$) CNBL. Uniform inflow wake deflections are overlaid with the dotted lines. }
    \label{fig:wake_centroid}
\end{figure}

Wake deflection, which is shown in \cref{fig:xy_wakes}, is influenced by the relative strength of Coriolis forces. 
Here, we measure the wake deflection by computing the centroid of the streamtube position, consistent with the analysis in \cref{ssec:results_uniform}. 
% The qualitative results do not change when considering the centroid of the velocity deficit field $\overline{\Delta u}$ at hub height or when considering the 3D wake, which is discussed in \cref{appx:streamtube_sensitivity}. 
The wake centroid evolution $y_c(x)$ is shown in \cref{fig:wake_centroid} for varying Rossby number with each inflow type. 
Because the hub height wind direction oscillates after the wind angle controller is turned off at the beginning of the concurrent simulations, the wind direction drifts up to \SI{1}{\degree} throughout the averaging time of the primary simulation. 
The advection due to the time-mean wind direction at hub height is removed by subtracting $x\tan(\winddir_{hub}) \approx \winddir_{hub} x$ from the streamtube centroid $y_c(x)$, where $\winddir_{hub}$ is given in \cref{tab:inflow_properties}. 

In all inflow conditions (uniform, TNBL, and CNBL), the amount of wake deflection is parametrically dependent on the relative strength of Coriolis forcing. 
The sign (clockwise or anti-clockwise) of the wake deflection is also dependent on the relative strength of Coriolis forcing for wakes in the ABL. 
This differs from the uniform inflow cases, which only deflect anti-clockwise in the northern hemisphere. 
Between $Ro = 500$ and $Ro = 250$, we observe a transition in wake deflection direction in the ABL. 
For $Ro = 500$, wakes in ABL inflow are deflected clockwise as viewed from above ($-y$ direction), while for $Ro \leq 250$, wakes are deflected anti-clockwise ($+y$ direction). 
While the transition $Ro$ between clockwise and anti-clockwise deflection will depend on ABL characteristics, such as $z_0$ and therefore wind shear, we expect that the qualitative trends presented here will apply to other neutral ABLs. 
Additionally, the wake deflection is not bounded above or below by the uniform inflow results. 
For example, in the CNBL at $Ro = 100$, the observed wake deflection is several times larger than the wake deflection in uniform inflow for $x/D \lesssim 6$.  

\begin{figure}
    \centering
    \includegraphics[width=\linewidth]{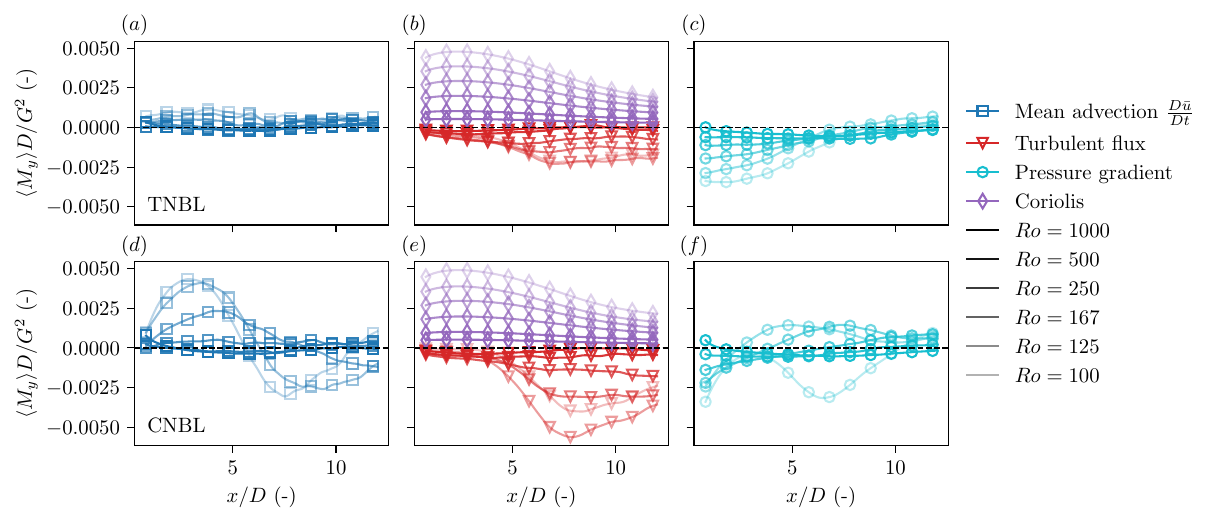}
    \caption{Streamtube-averaged momentum terms in the $y$-momentum balance for wakes in ($a$, $b$, $c$) TNBL inflow and ($d$, $e$, $f$) CNBL inflow. }
    \label{fig:abl_ymomentum}
\end{figure}

To understand why wakes are deflected increasingly anti-clockwise with decreasing Rossby number (increasing relative Coriolis forcing strength), we examine the lateral momentum budget terms $M_y$ averaged within the streamtube, shown in \cref{fig:abl_ymomentum}. 
In the TNBL, the mean advection term is of the same order of magnitude as in uniform inflow. 
The mean advection of the streamtube region is increasingly positive as the Rossby number decreases. 
Because the wake deflection is the integrated form of the mean lateral advection, the increasing positive advection term results in increasingly anti-clockwise wake deflection. 
The mean advection is the sum of the Coriolis, lateral pressure gradient, and turbulent flux forcing terms (subgrid forcing is negligible). 
As shown in \cref{fig:abl_ymomentum}($a$-$c$), the Coriolis, pressure gradient, and Reynolds stress terms are all dynamically active in the wake, but the relative importance of each term varies with streamwise location $x$. 

\begin{figure}
    \centering
    \includegraphics[width=0.9\linewidth]{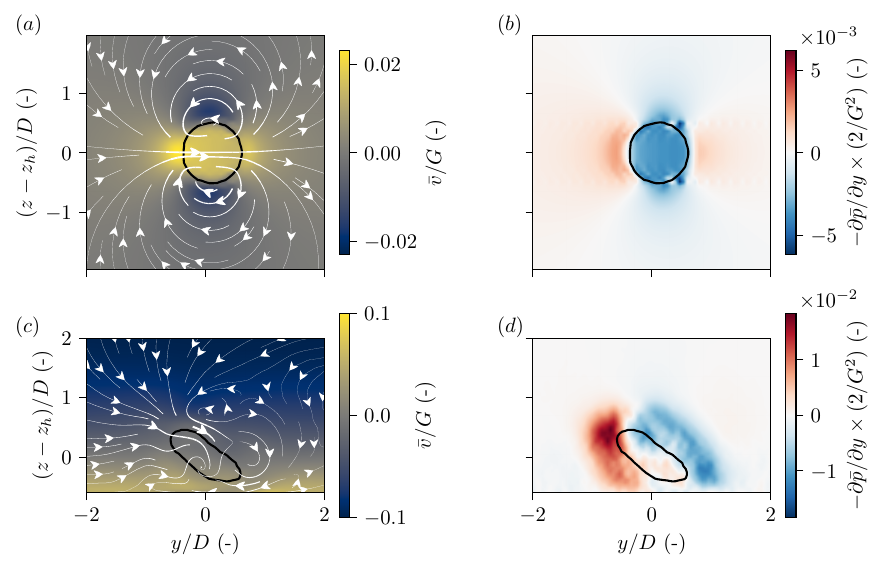}
    \caption{Wake cross sections at $x = 8D$ for ($a$, $b$) uniform inflow and ($c$, $d$) TNBL simulations at $Ro = 100$. ($a$, $c$) Contours of lateral velocity $\bar{v}$ are shown with streamlines of $\overline{\Delta v}$ and $\overline{\Delta w}$ overlaid. ($b$, $d$) Lateral pressure gradient fields in the wake. Note the different colorbar magnitudes. The streamtube boundary is shown by the black line. }
    \label{fig:cvp_formation}
\end{figure}

The near-wake dynamics are dominated by the direct Coriolis forcing. 
Similar to the uniform inflow simulations, the direct Coriolis forcing is again partially opposed by a lateral pressure gradient force. 
However, unlike in uniform inflow, the pressure gradient forcing term begins to decay almost immediately after the initial wake expansion. 
The decay of lateral pressure gradients in the TNBL occurs in conjunction with the lack of a coherent CVP formation in the TNBL turbine wake. 
A coherent CVP is not formed in the TNBL, as shown in the color contours of \cref{fig:cvp_formation}($c$), due to atmospheric turbulence. 
The lateral pressure gradients associated with the CVP formation in uniform inflow are obscured by turbulence in the TNBL wake. 
The lateral pressure gradient field primarily opposes the Reynolds stress divergence, similar to pressure gradients in turbulent jet flows \citep{pope_turbulent_2000}. 
The asymmetry in the pressure gradient forcing in \cref{fig:cvp_formation}($d$) is due to the Coriolis force. 
We note that a weak CVP can be observed in streamlines of the velocity deficit $\overline{\Delta v}$, $\overline{\Delta w}$ in the TNBL wake, seen in \cref{fig:cvp_formation}($c$). 
Even so, the structure of the full lateral velocity field $\bar{v}$, shown by the color contours, and pressure gradient field indicate that the effect of the CVP formation does not constitute the dominant mode for momentum transport. 

As the wake recovers, the direct Coriolis forcing term, which is proportional to the velocity deficit, begins to diminish. 
At the same time, the turbulent entrainment of lateral momentum increases. 
In the TNBL, we observe that the pressure gradient term responds to the combined Coriolis and pressure gradient forcing. 
For instance, the pressure gradient forcing flips sign in the wake region when the Reynolds stress divergence exceeds the direct Coriolis forcing. 
As a result, even in the far wake when the wake begins to recover and the direct Coriolis forcing decays, the wake continues to deflect in the TNBL, as quantified in \cref{fig:wake_centroid}. 
In contrast with the uniform inflow wakes, where the CVP formation drives the lateral pressure gradient forcing, turbulence primarily controls the lateral pressure gradient for wakes in the TNBL. 

Comparing the TNBL and CNBL simulations in \cref{fig:abl_ymomentum}, we see both similarities and differences in the lateral momentum budgets. 
For example, the direct Coriolis forcing term, which depends only on wake velocity and Rossby number, are nearly the same in the TNBL and CNBL. 
This is because the streamtube-averaged wake velocity is very similar between the TNBL and CNBL for all $Ro$, as shown in \cref{fig:wake_deficit}. 
However, the magnitude of the turbulent flux of lateral momentum into the wake is greater in the CNBL than in the TNBL, in general. 
The lateral momentum flux in the CNBL is generally larger than in the TNBL because the magnitude of wind direction shear is larger in the CNBL than in the TNBL for the same Rossby number.
The most substantial difference between the TNBL and CNBL lateral momentum budgets is the presence of oscillations in the lateral pressure gradients with wavelengths of several turbine diameters or greater. 
These oscillations are from gravity waves, as mentioned in \cref{ssec:results_streamwise}, and we will elaborate on the gravity wave behavior in \cref{ssec:results_waves}. 
Where waves are present ($Ro \lesssim 167$), oscillations due to gravity waves are the dominant contribution to the lateral pressure gradient forcing term. 
In the $Ro = 100$ and $Ro = 125$ simulations, the lateral pressure gradient decays rapidly in the near wake, which drastically increases the net lateral advection and therefore wake deflection in the CNBL inflow compared with the TNBL or uniform inflow. 
While the dynamics differ between the TNBL and CNBL due to free atmosphere stratification, we emphasize that the qualitative trends in wake deflection (\cref{fig:wake_centroid}) are the same: we observe parametrically varying wake deflection which can be clockwise or anti-clockwise, but is increasingly anti-clockwise as relative Coriolis forcing strength increases (Rossby number decreases). 

\begin{figure}
    \centering
    \includegraphics[width=0.9\linewidth]{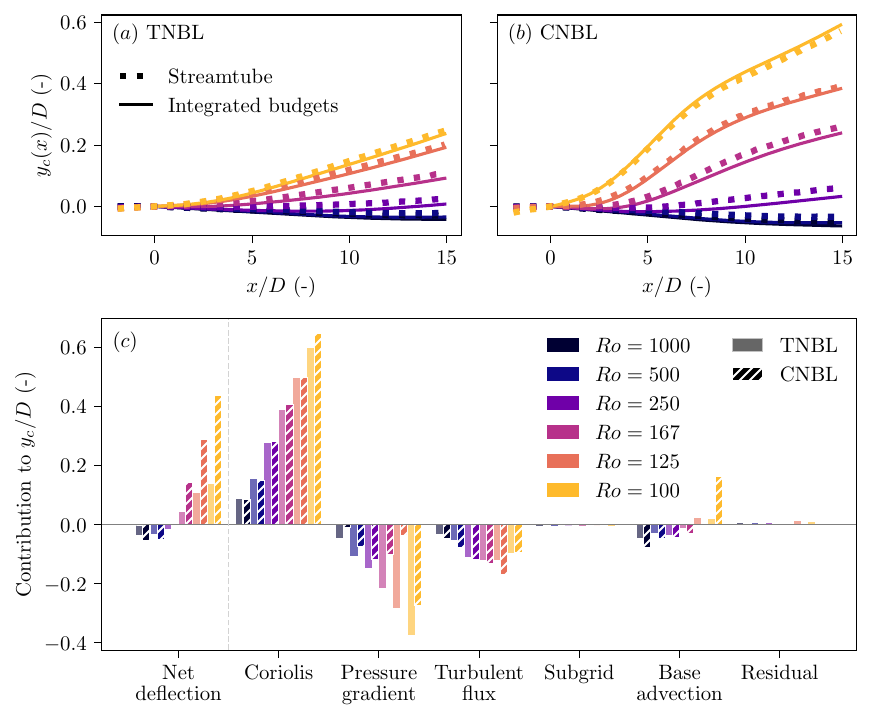}
    \caption{Integrated form of the lateral momentum budgets to reconstruct the streamtube centroid deflection in ($a$) TNBL inflow and ($b$) CNBL inflow. ($c$) Integrated contributions to the net streamtube deflection for each lateral forcing term at $x = 10D$. The net deflection, $y_c(x=10D)$, is equal to the sum of deflection contributions from all forcing terms to the right of the vertical dashed line. }
    \label{fig:integrated_ymom}
\end{figure}

The parametric dependence of the lateral momentum budget terms on the Rossby number has a direct connection to the wake deflection. 
In particular, because the momentum budget terms represent a forcing (and therefore acceleration) on the flow, integrating the lateral momentum budgets twice in time first yields lateral velocity, then wake deflection. 
Additionally, using the streamtube-averaged velocity $\langle \bar{u}\rangle$ to advect the flow, time integration can be transformed into spatial integration in the streamwise direction $x$. 
A mathematical derivation including the assumptions and transformations which connect the wake deflection to the forcing terms $\bar{f}_y$ in the lateral RANS  budget (terms II-V in \cref{eq:V_RANS}) are given in \cref{appx:integration}. 
The final result is
\begin{equation}
    \label{eq:yc_integration}
    y_c(x) = 
    \sum 
    \underbrace{
    \int_{x_0}^{x} \frac{1}{\langle \bar{u} \rangle (x')}
    \int_{x_0}^{x'} \frac{1}{\langle \bar{u} \rangle (x'')} 
    \langle \bar{f}_y(x'', y, z) \rangle \,dx'' \,dx'
    }_\text{Individual forcing contributions}
    + \underbrace{
    \int_{x_0}^x \frac{\langle \bar{v} \rangle (x_0)}{\langle \bar{u} \rangle (x')}
    \,dx' 
    }_\text{Base Advection}. 
\end{equation}
As a result of the integration, the observed streamtube deflection (\cref{fig:wake_centroid}) can be reconstructed from the streamtube-averaged momentum budgets, as shown in \cref{fig:integrated_ymom}($a$-$b$). 
Overall, the integrated budgets reconstruct the observed streamtube lateral deflection well, quantitatively capturing the streamtube deflection amount. 

To extend this analysis, we represent the integrated contribution of each lateral forcing term as an equivalent wake deflection amount at $x = 10D$, shown in \cref{fig:integrated_ymom}($c$). 
Some forces, such as direct Coriolis forcing, cause anti-clockwise ($+y$ direction) wake deflection in the northern hemisphere. 
Other forces, such as the turbulent flux of lateral momentum deflect wakes clockwise. 
Importantly, the turbulent flux of lateral momentum does not increase in magnitude as fast as the direct Coriolis forcing, resulting in increasingly anti-clockwise wake deflections with decreasing $Ro$. 
The lateral pressure gradient force generally opposes the Coriolis force, although in the CNBL, the integrated magnitude of the pressure gradients is strongly affected by gravity waves. 
For example, at $Ro = 125$ with the integration bounds $x \in [0, 10]D$, the pressure gradient contribution to the wake deflection is nearly zero (see \cref{fig:abl_ymomentum}($f$)). 
While the net deflection and direct Coriolis forcing contribution monotonically increase with increasing relative Coriolis forcing (decreasing $Ro$), some forces, such as the lateral pressure gradients and turbulent fluxes, vary non-monotonically as a result of structural changes in the ABL with decreasing $Ro$ such as decreasing inflow $TI$. 
Finally, the base advection term represents the non-linear interaction between the non-zero lateral velocity at the rotor, which is due to the wind direction drift in the ABL and to Coriolis effects in the induction region of the wind turbine \citep{gadde_effect_2019}, with the wake velocity deficit. 
As \cref{fig:integrated_ymom}($c$) shows, multiple lateral forcing terms have a leading-order importance on the wake deflection of a wake in the ABL, and each forcing term has a different parametric dependence on the relative strength of Coriolis forcing. 

The integrated lateral momentum budgets link the momentum budgets with the observed wake deflections. 
Due to the transformation from time coordinates to spatial coordinates through the streamwise velocity $\langle \bar{u}(x) \rangle$, the lateral forcing (acceleration) which occurs in the near-wake has a stronger effect on the wake deflection than forcing in the far-wake. 
This is because the wake velocity in the near-wake region is slower than in the far-wake, and therefore the residence time of the near-wake is longer than in the far-wake. 
Notably, the dominant physics in the near-wake region is the anti-clockwise Coriolis forcing, shown in \cref{fig:abl_ymomentum}. 
Assuming a linearized advection velocity $\langle \bar{u}(x) \rangle \approx u_{hub}$ fails to reconstruct the wake deflection from integrating the momentum budget and persistently underpredicts the wake deflection due to Coriolis effects. 
In summary, through a streamtube-averaged analysis of the wake momentum budget, we parse the importance of each ABL forcing term on the net wake deflection to explain the parametric variation of wake deflection on relative Coriolis forcing strength.

\subsection{Gravity waves induced by stand-alone turbines}
\label{ssec:results_waves}

For CNBLs with shallow boundary layer heights, we observe oscillations in the time-averaged wake velocities, pressure fields, and momentum budgets. 
The pressure oscillations in the streamwise and lateral momentum budgets are indicative of atmospheric gravity wave-like behavior due the presence of the lone wind turbine for low Rossby numbers. 
% Background may need to be moved to e.g., introduction
In the context of wind energy, gravity waves are traditionally studied as a wind farm-scale phenomenon where the flow deceleration accompanying wind turbine power extraction perturbs the capping inversion layer, which is a region of strong stable stratification that can act as a lid on the ABL \citep{allaerts_boundary-layer_2017}. 
The perturbed capping inversion can excite gravity waves, which propagate laterally and vertically into the stably stratified free atmosphere. 
Gravity waves have been simulated in LES for semi-infinite (finite only in the streamwise direction) and finite wind farms in both conventionally neutral and stable ABLs \cite[\eg][]{allaerts_boundary-layer_2017, wu_flow_2017, allaerts_gravity_2018, lanzilao_parametric_2024}. 

Gravity waves non-locally redistribute kinetic energy throughout large wind farms due to pressure oscillations. 
The non-local redistribution of energy typically creates a farm-scale adverse pressure gradient in the leading rows of a wind farm, further decelerating the incoming freestream flow, with a favorable pressure gradient further downwind (typically inside the wind farm) which accelerates the flow \citep{lanzilao_parametric_2024}. 
Often, studies will compare the effects of atmospheric gravity waves on a wind farm to the performance of a single, stand-alone turbine \citep{lanzilao_parametric_2024, sanchez_gomez_investigating_2023}. 
This assumes either that the stand-alone turbine does not excite gravity waves, or that the gravity wave effects are small relative to the turbine-specific dynamics such as the induction region \citep{sanchez_gomez_investigating_2023}. 
Previous studies which use an isolated turbine as a baseline for comparing the energy redistribution as a result of gravity waves have not observed a stand-alone turbine inducing atmospheric gravity waves.

\begin{figure}
    \centering
    \includegraphics[width=\linewidth]{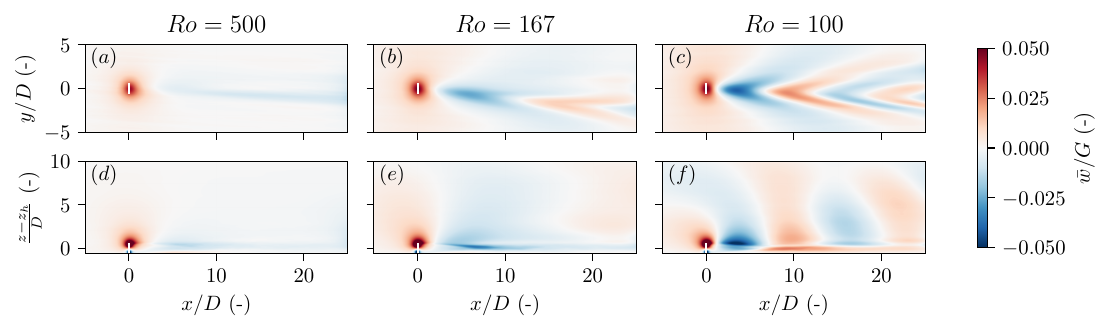}
    \caption{Mean vertical velocity fields $\bar{w}$ for ($a$, $d$) $Ro = 500$, ($b$, $e$) $Ro = 167$, and ($c$, $f$) $Ro = 100$ in CNBL inflow. Slices are shown for ($a$-$c$) a horizontal plane at $z = z_h + D$ and ($d$-$f$) a vertical plane through the rotor center ($y=0$). 
    Gravity waves propagate with the background flow velocity, which is misaligned with the $x$-direction due to wind direction shear.
    As a result, the waves shown in ($b$) and ($c$) are advected in the $-y$ direction.}
    \label{fig:waves_xy}
\end{figure}

Atmospheric gravity waves are observed in the present low Rossby number CNBL simulations where the capping inversion layer directly interacts with the upper tip of the rotor. 
No wave-like behavior is observed in TNBL simulations because the domain does not have a stably stratified free atmosphere. 
The highest CNBL height in which gravity waves are observed occurs at $h = \SI{272}{\meter}$ ($Ro = 167$), compared to the rotor tip height of \SI{270}{\meter}. 
Note that the definition of the boundary layer height used here is based off of the turbulent shear stress and not the inversion layer location, but these heights are similar and inter-dependent. 
Vertical velocity perturbations associated with atmospheric gravity waves are shown in \cref{fig:waves_xy} for select Rossby numbers.  

The pressure gradient forcing from the induced gravity waves is dynamically important to the streamwise and lateral momentum budgets. 
In the streamwise momentum budget, the presence of gravity waves creates a favorable pressure gradient region directly behind the rotor which creates a region of accelerated wake recovery, followed by a region of adverse pressure gradient. 
The oscillations in the pressure field decrease in wavelength as the geostrophic wind speed decreases, as predicted by linear theory ($\lambda_x = 2\pi G_x/N$, where $N = \sqrt{g \Gamma /\theta_0}$ is the \BV frequency).  
% show table? 
Although the momentum budgets show a favorable pressure gradient in the region of a downwind turbine, the addition of a second, waked turbine would likely alter the gravity wave development as gravity wave induction is inherently non-local. 
Therefore, future simulations of multiple turbines in low Rossby number CNBLs should be conducted to study the coupled effects of gravity waves on turbine-scale wake interactions. 

In the lateral momentum budget, the presence of gravity waves strongly affects the lateral pressure gradient forcing. 
In the uniform inflow and TNBL simulations, the lateral pressure gradient initially opposes the dominant Coriolis forcing term in the near wake region. 
The opposition of the pressure gradient forcing to the Coriolis forcing limits the net advection, and by extension, the wake deflection. 
In the CNBL where gravity waves are present, the dominant physics governing the pressure field in the near-wake region is no longer the Coriolis forcing. 
Therefore, the Coriolis forcing term is unbalanced in the near wake, resulting in a larger positive lateral advection (and therefore wake deflection) in the CNBL than in the TNBL or in uniform inflow. 

We hypothesize that inception of gravity waves due to a stand-alone turbine have not previously appeared in the literature because turbines historically have not operated in the upper extent of the ABL. 
However, new offshore wind installations have recently exceeded the height of the \IEA reference turbine in Europe and China. 
% cite e.g., https://www.euronews.com/green/2023/09/07/world-record-wind-turbine-generates-enough-energy-in-a-day-to-power-170000-homes
In addition, stable boundary layers such as the canonical GABLS case \citep{beare_intercomparison_2006} exhibit boundary layer heights that are suppressed in height by strong stratification and typically reach 100-500~m in height \citep{stull_introduction_1988}. 
For large onshore or offshore turbines located where stable boundary layers are common, a large turbine may also directly interact with the free atmosphere. % cite papers? 
Therefore, for the next generation of wind turbines, interactions between the rotor, wakes, and the free atmosphere should be explicitly investigated.

\subsection{Contextualizing Coriolis-driven wake deflection with wind farm flow control}
\label{ssec:results_implications}

Atmospheric conditions typically change on time scales of $\mathcal{O}(\SI{1}{\hour})$ \citep{stull_introduction_1988}. 
A time averaging period of one inertial period $2\pi/f_c$ is used in the ABL simulations presented in this study to ensure that the reported dynamics are not affected by a partial period of an inertial oscillation. 
This time averaging period is much longer than the response time of the ABL. 
In other words, the ABL is rarely quasi-stationary for a whole inertial oscillation period. 
In this section, we compare the variation across Rossby numbers to the inherent variation in turbulence over one-hour averages. 

\begin{figure}
    \centering
    \includegraphics[width=\linewidth]{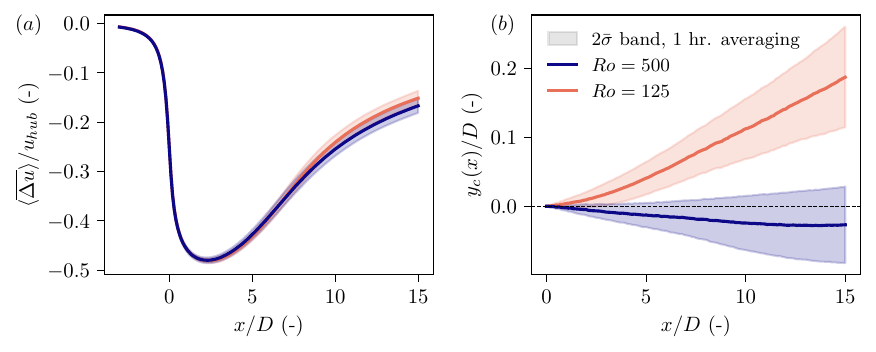}
    \caption{Variation in ($a$) wake recovery and ($b$) wake deflection for two Rossby numbers in TNBL inflow, showing a distribution from one-hour time-averaged flow fields. }
    \label{fig:window_1hr}
\end{figure}

To assign an interval of uncertainty to the reported wake statistics, we compare the full time average over one inertial period with independent one-hour time averages. 
The standard error of the mean $\bar{\sigma}$ is computed from the standard deviation of the ensemble of one-hour averages. 
A 2$\bar{\sigma}$ band around the sample mean wake deflection and wake recovery is shown in \cref{fig:window_1hr} for two Rossby numbers in TNBL inflow. 
The Rossby numbers $Ro = 125$ and $Ro = 500$ approximately correspond with cut-in and rated wind speed for the \IEA reference turbine. 

The one-hour time-averaged metrics highlight the internal variability of turbulence in the ABL. 
For the streamwise velocity deficit shown in \cref{fig:window_1hr}($a$), the confidence intervals between the two Rossby number values overlap. 
This indicates that the variability in wake recovery due to stochastic turbulence in the TNBL is greater than the variability in wake recovery due to Coriolis effects. 
% This indicates that for wake recovery, parametric variance to the strength of Coriolis forcing through $Ro$ is much smaller than variance due to chaotic ABL turbulence. 
That is, for the range of Rossby numbers studied here, we find no statistically significant impact of Coriolis effects on wake recovery of a free-standing turbine, relative to the internal variability of the ABL, for time scales relevant to the ABL or to wind turbine control. 

% Variance in wake deflection between Rossby numbers is greater than the variance from ABL turbulence. 
In contrast, the impact of relative Coriolis forcing strength on wake deflection is statistically significant, as shown in \cref{fig:window_1hr}($b$). 
For the Rossby number values relevant to the \IEA reference turbine, the difference in wake deflection due to Coriolis forcing exceeds the variance due to ABL turbulence over one-hour time scales. 
Additionally, the difference in wake deflection between Rossby numbers increases with increasing distance downwind. 
It is also interesting to note that for the weakest relative Coriolis forcing ($Ro = 500$), the $2\bar{\sigma}$ envelope of variability includes zero and anti-clockwise wake deflections. 
Previous literature has focused on shorter time averaging lengths than presented here. 
Alongside variation in the ABL conditions, the spread of reported wake deflection may be partially attributed to variability of ABL turbulence over relatively short time scales. 

% Quick note about yaw misalignment
Wake steering, or the intentional deflection of wind turbine wakes induced through yaw-misalignment of the rotor, has garnered attention as one way to increase collective power production of wind turbine arrays \citep{fleming_initial_2019, howland_collective_2022}. 
Previous work has focused on the combined effects of Coriolis forcing and wake steering, which both can act to deflect wakes in the ABL. 
In LES of a five-turbine array at $Ro = 1005$, \citet{nouri_coriolis_2020} used wake steering to increase aggregate farm power production. 
They found that power production gains for positive yaw-misalignment angles increased over negative yaw-misalignments when Coriolis forces were present in the inflow \citep{nouri_coriolis_2020}. 
Here, positive yaw-misalignments are anti-clockwise turbine rotations about the vertical axis, which induce clockwise wake deflection. 
Similarly, \citet{wei_parametric_2023} recommend positive yaw-misalignments for fully-waked conditions in the northern hemisphere. 

While we focus on wakes of yaw-aligned turbines in this study, we can quantify the magnitude of equivalent yaw-misalignment which would equal the wake deflection observed due to Coriolis forcing alone. 
To predict the wake deflection due to yaw misalignment, we use the model proposed in \citet{shapiro_modelling_2018} and extended in \citet{heck_modelling_2023} using a wake spreading rate of $k_w = 0.083$ and the same turbine thrust coefficient $C_T' = 1.33$. 
We find that for $Ro = 125$, the equivalent yaw-misalignment for the observed wake deflection at $x = 10D$ in the TNBL is approximately $-8^\circ$, while in the CNBL it is approximately $-17^\circ$. 
The equivalent yaw-misalignment angles are negative because for strong relative Coriolis forcing (low Rossby numbers), we observe that the Coriolis force deflects wind turbine wakes anti-clockwise.
Therefore, for the Rossby number ($Ro = 125$) and ABL conditions simulated here, positive yaw misalignment and Coriolis-based wake deflections would be in opposition and would nearly cancel out, while negative yaw misalignment will enhance the inherent Coriolis-based anti-clockwise deflection.
Future research should study the interactions of relative Coriolis forcing strength with wake steering control. 

Previous studies that investigated Coriolis effects on wind turbine wakes have focused on turbines about half the size of modern offshore wind turbines.  % cite papers? 
Studying a rotor twice as large in diameter halves the relevant Rossby numbers for that turbine. 
For example, the NREL~\SI{5}{\mega\watt} turbine operating at rated wind speed sees a Rossby number $Ro \approx 880$. 
Our results suggest that in this operating regime, wakes for the NREL~\SI{5}{\mega\watt} reference turbine would deflect clockwise \citep{van_der_laan_why_2017}. 
However, as turbines become larger, Coriolis forces become more dynamically important, relative to the other terms in the lateral momentum balance. 
The next generation of turbines with rotors greater than \SI{200}{\meter} in diameter will increasingly operate in regimes where the dynamics studied here are relevant, and wakes will be deflected increasingly anti-clockwise for lower Rossby numbers.

\section{Summary and conclusions}
\label{sec:conclusion}
% discussion section? 

% - Introduce the field.
	% - Introduce the problem.
	% - What is the knowledge gap?
	% - What are you doing about it?
	% - How are you doing it?
	% - Why is it important?
 
The presence of Coriolis forces in the ABL affects wind turbine wake dynamics. 
% by altering the structure of the ABL as well as through direct Coriolis forcing. 
Previous investigations have studied the interaction between turbine wakes and Coriolis forces for a relatively small range of rotor diameter-based Rossby numbers ($Ro = G/(f_c D)$). 
As wind turbine rotors increase in size, the wake structures that they create in the ABL also grow, resulting in more dynamically active Coriolis forcing. 
We explore a Rossby number range relevant to the next generation of offshore wind turbines using the \IEA turbine for reference. 

In this study, large eddy simulations of a single actuator disk modeled wind turbine are used to explore the parametric effects of Coriolis forces on wind turbine wake dynamics in uniform and boundary layer inflow. 
Using a streamtube control volume, the lateral momentum budget is linked with observations of wake deflection. 
In uniform inflow conditions, the Coriolis force causes wakes to deflect anti-clockwise. 
When scaled by the Rossby number, wake dynamics in uniform inflow are shown to be self-similar, and non-negligible lateral pressure gradients partially oppose the direct Coriolis forcing. 
Lateral pressure gradients arise due to the formation of a counter-rotating vortex pair in uniform inflow and limit the wake deflection magnitude. 
For uniform inflow, neglecting the lateral pressure gradients leads to an order-of-magnitude overprediction in wake deflection. 

For TNBL and CNBL inflow, wake deflection also monotonically increases with decreasing Rossby number, becoming increasingly anti-clockwise. 
At high Rossby numbers (weak Coriolis forcing strength relative to inertial forces), wakes are deflected clockwise due to the effects of wind speed and direction shear that stem from Coriolis effects on the background ABL flow, in agreement with previous studies that have investigated high Rossby number regimes. 
At low Rossby numbers, wakes are deflected anti-clockwise. 
In the simulations presented here, we observe a transition between clockwise wake deflection to anti-clockwise wake deflection between $Ro = 500$ and $Ro = 250$. 
The transition point will depend on the ABL inflow conditions, such as the surface roughness, and wind turbine design and control strategy (\eg thrust coefficient), which should be studied in future work. 
Unlike in uniform inflow, for ABL simulations, the dynamically active terms in the lateral momentum budget change as a function of streamwise position. 
Additionally, the lateral pressure gradient contribution is non-negligible and changes with the dominant wake dynamics. 
In the TNBL, as well as in the CNBL when gravity waves are not present, the sign of the pressure gradient forcing term changes when the Reynolds stress divergence overtakes the Coriolis forcing term. 
In the CNBL, gravity waves are triggered when the boundary layer height interacts with the upper tip of the rotor area. 
Gravity waves considerably alter the dynamics of the lateral advection; in the present simulations, the presence of gravity waves increase the wake deflection in the CNBL compared to the TNBL or uniform inflow. 
While shallow CNBLs may be relatively uncommon in the environment, it is certainly possible that for stable boundary layers, large, modern wind turbines could directly interact with the free atmosphere and trigger gravity waves. 
We recommend that future work studies the effects of gravity waves induced by stand-alone turbines. 

This study is focused on neutrally stratified ABLs to limit the complexity of the parameter space and focus on Coriolis effects on wind turbine wakes. 
Thermal stratification, which, similar to Coriolis effects, alters the structure of wind shear and turbulence in the ABL, also affects wake dynamics \citep[\eg][]{abkar_influence_2015}. 
Further, only the vertical component of Earth's rotation is considered in this work, while the horizontal component has been shown to affect ABL development and vertical transport of Reynolds stresses \citep{howland_influence_2020}.  
Future work should consider the joint effects of Coriolis and buoyancy effects on wake dynamics. 
Additionally, we only simulate one surface roughness value representative of offshore wind conditions. 
Finally, future work should investigate the effects of Coriolis forcing on wind farm wake dynamics to explore the effects of wake superposition in varying Rossby number regimes. 
Understanding these effects will improve future wake models and influence wind farm design and control.

\section*{Acknowledgements}
The authors acknowledge funding from the National Science Foundation (Fluid Dynamics program, grant number FD-2226053).
In addition, K.S.H. acknowledges funding through a National Science Foundation Graduate Research Fellowship under grant no. DGE-2141064. 
% M.F.H. acknowledges partial support from the MIT Energy Initiative.
Simulations were performed on Stampede2 and Stampede3 supercomputers under the NSF ACCESS project ATM170028.

\section*{Declaration of Interests}
The authors report no conflict of interest.

% ======= Appendices begin here =======
% \vfill \break

\begin{appendix}
\section{Dependence on time averaging window}
\label{appx:time_averaging}

In the absence of friction, the free atmosphere behaves like an undamped harmonic oscillator. 
Perturbations from the geostrophic balance will cause inertial oscillations, which are periodic changes in the ABL structure that occur on time scales of $\approx 2\pi/f_c$ \citep{blackadar_boundary_1957, stull_introduction_1988}. 
In LES, inertial oscillations are introduced at the initialization of the ABL and slowly decay due to damping from the ABL. 
However, we still observe changes of the wind speed of about $\pm 1\%$ and changes in wind direction of $\pm 0.5^\circ$ over the duration of one inertial period. 
Therefore, we choose to time-average wake statistics over a time interval $T = 2\pi/f_c \approx \SI{17}{\hour}$ to avoid biasing results with statistics which contain a partial period of inertial oscillation. 

The time averaging duration can be increased to any integer multiple of the inertial oscillation period. 
To check if our results are sensitive to the time averaging period, we also perform a simulation for the $Ro = 125$ case in TNBL inflow over two inertial periods $4\pi / f_c \approx \SI{34}{\hour}$. 
The comparison of wake deflections averaged over one and two inertial periods is shown in \cref{fig:appx_yc_sensitivity}($a$). 
We find that the quantitative wake deflection only change by a maximum of 4\% at a location of $x = 5D$ when comparing the first inertial oscillation period to the second. 
Therefore, we find that a single inertial period of time averaging is sufficient for the statistics reported here. 
Additionally, we find that the wake recovery is less sensitive than the wake deflection to the averaging time, and we also find no significant difference in the residuals of the momentum budgets when averaging over a double period of inertial oscillations. 

\begin{figure}
    \centering
    \includegraphics[width=\linewidth]{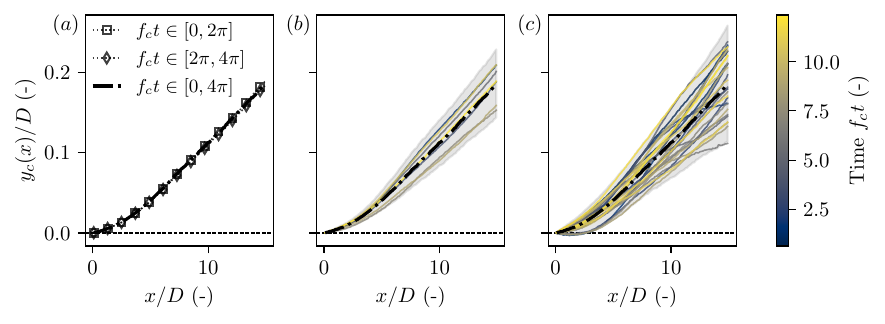}
    \caption{Sensitivity to time averaging for wake deflections at $Ro = 125$ in TNBL inflow. The dash-dotted line is time averaged over a double period of inertial oscillations. ($a$) Averaging over one inertial period ($\approx \SI{17}{\hour}$) removes all scatter in the wake centroid data. ($b$) Wake deflections for 4-hour time averages. ($c$) Wake deflections for 1-hour time averages. In ($b$) and ($c$), the gray shaded region represents a $2\bar{\sigma}$ interval.}
    \label{fig:appx_time_avg}
\end{figure}

In \cref{ssec:results_implications}, one-hour time averages are used to show a band of possible wake deflections on time scales relevant to ABL dynamics and wind farm flow control. 
This band is computed with a width of twice the standard deviation $\bar{\sigma}$ of the ensemble of wake deflections as a function of downstream distance. 
Wake deflections measured in non-overlapping 1-hour and 4-hour time averages are shown in \cref{fig:appx_time_avg}($b$) and ($c$), respectively. 
As the averaging time increases, the variation in the observed range of deflections narrows around the mean value, which averaged over a double inertial period. 
The time in which the averaging window ends is denoted by the colorbar for \cref{fig:appx_time_avg}($b$, $c$), where $t = 0$ corresponds to the time that the wake has fully developed and time averaging begins. 
The spread in colors between instances of wake deflection shows that the wake deflection does not drift, but rather it changes stochastically with ABL turbulence.

\section{Streamtube analysis sensitivity}
\label{appx:streamtube_sensitivity}

\begin{figure}
    \centering
    \includegraphics[width=\linewidth]{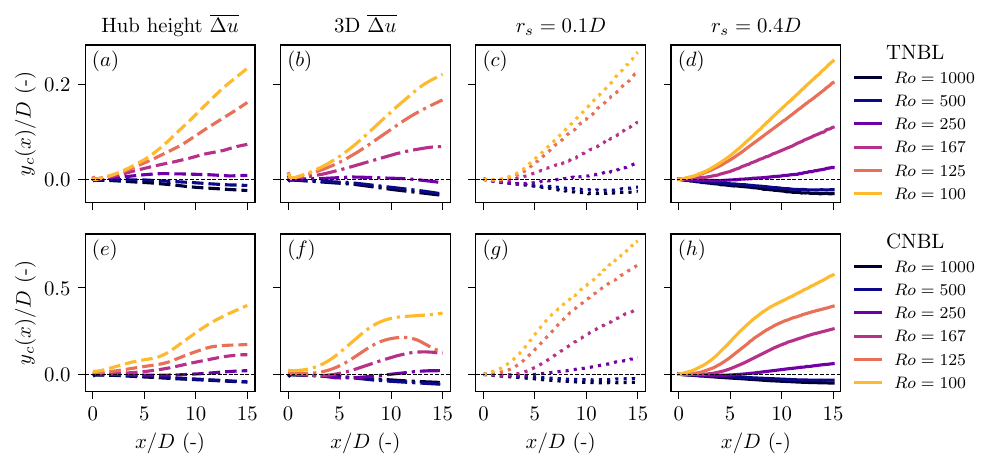}
    \caption{Wake centroid deflection in ($a$-$d$) TNBL inflow and ($e$-$h$) CNBL inflow for various definitions of the wake center $y_c$. The wake center is given by ($a$, $e$) hub-height centroid of $\overline{\Delta u}$ given by \cref{eq:centroid_hub}, ($b$, $f$) three-dimensional centroid of $\overline{\Delta u}$ given by \cref{eq:centroid_3D}, ($c$, $g$) streamtube centroid seeded at $r_s = 0.1D$, and ($d$, $h$) streamtube centroid seeded at $r_s = 0.4D$, which also is shown in \cref{fig:wake_centroid}.}
    \label{fig:appx_yc_sensitivity}
\end{figure}

% Introduce the appendix: what are the goals here
In this appendix, we compare the streamtube analysis of wake deflection to other wake deflection metrics. 
As discussed in \cref{ssec:streamtubes}, using a streamtube as an analog for the wake provides a physical and robust interpretation of the wake evolution. 
Throughout this paper, wake deflection $y_c$ is defined as the centroid of the streamtube in the $yz$-plane (\cref{fig:streamtubes}($b$)). 
Other definitions of the wake deflection include the centroid of the wake velocity deficit at hub height, computed as 
\begin{equation}
    \label{eq:centroid_hub}
    y_c(x) = \frac
    {\int y\overline{\Delta u} \,dy}
    {\int \overline{\Delta u} \, dy}, 
\end{equation}
or the centroid of the wake computed from a $yz$-cross-section of the wake \citep[\eg][]{howland_wake_2016}
\begin{equation}
    \label{eq:centroid_3D}
    y_c(x) = \frac
    {\iint y\overline{\Delta u} \,dy \,dz}
    {\iint \overline{\Delta u} \, dy \, dz}. 
\end{equation}
Various definitions of the wake deflection are shown in \cref{fig:appx_yc_sensitivity}. 
The definition of $y_c$ given by \cref{eq:centroid_hub} is shown in \cref{fig:appx_yc_sensitivity}($a$, $e$) while the definition of $y_c$ given by \cref{eq:centroid_3D} is shown in \cref{fig:appx_yc_sensitivity}($b$, $f$). 

For all definitions of the wake deflection $y_c$, we observe that decreasing the Rossby number (increasing the relative importance of Coriolis forces) results in increasingly anti-clockwise wake deflections. 
Additionally, we observe a split between low Rossby number, Coriolis-dominated anti-clockwise wake deflections and high Rossby number, turbulence-dominated wake deflections regardless of the definition of $y_c$ used. 
For all cases, the transitional Rossby number is between $Ro = 500$ and $Ro = 250$. 

\begin{figure}
    \centering
    \includegraphics[width=0.9\linewidth]{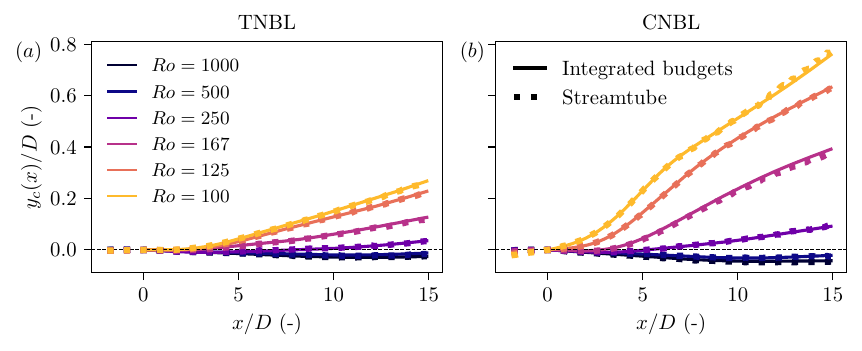}
    \caption{Same as \cref{fig:integrated_ymom}($a$, $b$) except using $r_s=0.1D$ rather than $r_s=0.4D$. Integrated budgets match the streamtube position more closely for the smaller streamtube radius. Dotted lines show the streamtube deflection for the smaller streamtube, and the dash-dotted lines show the integrated form of the lateral momentum budget.}
    \label{fig:appx_integrated_ymom}
\end{figure}

% Final streamtube plot: show reconstruction of budgets to contributions of y_c
We connect the observed wake deflections across ABL regimes with the lateral momentum budget in \cref{ssec:results_lateral}. 
Integrating the lateral momentum budget recovers the lateral position (streamtube deflection), as we show in \cref{appx:integration}. 
However, the reconstruction of the streamtube deflection \cref{fig:integrated_ymom}($a$, $b$) does not exactly match the streamtube position. 
This is because the streamwise velocity $\langle \bar{u} \rangle$ is used to transform temporal integration into spatial integration in $x$, but the streamwise velocity $\bar{u}$ varies within the streamtube due to wind shear in the ABL and the presence of the wake. 
% The derivation and details of this transformation, and the assumptions made in the integration reconstruction, are explained in \cref{appx:integration}. 
Velocity shear has a smaller effect on the integrated reconstruction of the wake deflection for smaller streamtubes, as explained by the dispersive advection in the derivation of the integral reconstruction (\cref{appx:integration}). 
The same analysis in \cref{fig:integrated_ymom}($a$, $b$) is shown in \cref{fig:appx_integrated_ymom} for a streamtube seeded at $r_s = 0.1D$, showing excellent agreement between the streamtube deflection and the integration reconstruction from the momentum budgets. 
In other words, the error incurred in the integration of the momentum budget is primarily due to the spatial averaging of $\langle \bar{u} \rangle$ and not because of a residual in the lateral momentum budget from LES. 

\section{Derivation of the integration of momentum budgets}
\label{appx:integration}

The goal of this section is to derive the transformation between the streamtube-averaged momentum budget equations and the observations of streamtube-averaged velocity $\langle \bar{u} \rangle (x)$ and streamtube deflection $y_c(x)$. 
We take a Lagrangian perspective of fluid motion which follows an infinitesimally small parcel along a streamline trajectory seeded at the labeling position $a_j$ at $\tau = 0$, where $\tau$ is the Lagrangian time. 
Newton's second law states that the sum of forces on the material volume representing the fluid parcel is equal to the change in its linear momentum. 
Written in terms of force per unit mass ($f_i$) as is consistent with the other equations in this paper, Newton's second law is given by 
\begin{equation}
    \label{eq:appx_newton_2}
    \frac{\partial u_{p, i}}{\partial \tau} 
    = \sum f_i(r_{p, i}(\tau), \tau), 
\end{equation}
where we use the subscript $p$ in $u_{p, i}(\tau) = u_i(\tau; a_j)$ to represent the velocity of the parcel seeded at the labeling parameter $a_j$. 
As in previous equations in this study, the subscript $i$ ($j$) denotes the direction of a tensor, such as the position and velocity vectors.
In general, the force on the parcel at time $\tau$ depends on the position of the parcel $r_{p, i}(\tau) = r_i(\tau; a_j)$. 
The velocity of the parcel is the time derivative of the parcel position such that 
\begin{equation}
    \label{eq:appx_pos_vel}
    u_{p, i}(\tau) = \frac{\partial r_{p, i}}{\partial \tau}. 
\end{equation}

To solve \cref{eq:appx_newton_2} from an initial position $a_j$ and initial velocity $u_{p, i}(0)$, we need information about the right hand side (RHS) forcing terms. 
In this analysis, we assume that the field $f_i(x_j)$ is statistically steady and known \textit{a priori} (\ie time-averaged fields from turbulent ABL flow). This is because our aim is to reconstruct the streamline trajectory $r_{p, i}(\tau)$ from the forcing field $f_i$, rather than write a functional form for the RHS forcing terms. 
When given the RHS forcing terms which are provided in \cref{eq:U_RANS} and \cref{eq:V_RANS} for the $x$ and $y$ directions, respectively, \cref{eq:appx_newton_2} and \cref{eq:appx_pos_vel} can be numerically integrated forward in time to solve for the streamline velocity and position. 
A change of variables from temporal integration to spatial integration is given by \cref{eq:appx_pos_vel} in the $i=1$ direction such that $d\tau = (u_p(\tau))^{-1} \,dx$ as long as the streamwise parcel
velocity $u_p = u_{p, 1}$ never changes sign, which is true for the flows described in this study. 
This change of variables also indicates that there exists a non-linear transformation $\tau \to x$, where $x$ is the streamwise coordinate in an Eulerian coordinate system. 
Under the spatiotemporal transformation $\tau \to x$, the position of the streamline can be written $r_{p, i}(\tau) = (x(\tau), y_p(x(\tau)), z_p(x(\tau))$, where the functions $y_p(x)$ and $z_p(x)$ are yet unknown. 
Note that $x_p(x(\tau)) = x$ because there is no coordinate transformation between the Lagrangian and Eulerian frames of reference.  
Then, an expression for the streamline velocity can be written by combining \cref{eq:appx_newton_2} and \cref{eq:appx_pos_vel} and applying the change of variables $\tau \to x$ such that
\begin{equation}
    \label{eq:appx_ux_diff_full}
    u_p(x) \frac{\partial u_{p, i}(x)}{\partial x}
    = 
    \sum f_i(x, y_p(x), z_p(x)). 
\end{equation}

To connect the integrated RHS forcing terms with the streamline position (and later, with the streamtube position), we can make \cref{eq:appx_ux_diff_full} an explicit expression for $u_{p, i}$ if the forcing on the parcel along the path of the streamline is also known \textit{a priori}. 
That is, by assuming a known streamline path $y_p(x) = \tilde{y}_p(x)$, $z_p(x) = \tilde{z}_p(x)$ to evaluate the forcing terms $f_i$, we can reconstruct the velocity at the points along the streamline path by slightly modifying \cref{eq:appx_ux_diff_full} such that 
\begin{equation}
    \label{eq:appx_ux_diff}
    u_p(x) \frac{\partial u_{p, i}(x)}{\partial x}
    = 
    \sum f_i(x, \tilde{y}_p(x), \tilde{z}_p(x)). 
\end{equation}
Integrating \cref{eq:appx_ux_diff} from $x_0 = x_p(\tau=0)$ to $x$ yields an expression for the streamline velocity: 
% If we make the additional assumption that we know the forces along the streamline trajectory, then we can write an explicit expression for the particle velocity using the change of variables as 
% 
% \begin{equation}
%     \label{eq:appx_ux_diff}
%     u_p(x) \frac{\partial u_{p, i}(x)}{\partial x}
%     = 
%     \sum f_i(x, \tilde{y}_p(x), \tilde{z}_p(x)). 
% \end{equation}
% % 
% Here $\tilde{y}_p$ and $\tilde{z}_p$ are known \textit{a priori}, for example by constructing a streamline $dx/\bar{u} = dy/\bar{v} = dz/\bar{w}$, where ${u}_i = ({u}, {v}, {w})$ is the Eulerian velocity field as defined in \cref{ssec:RANS}. 
% Integration in the $x$ direction from $x_0 = x(\tau=0)$ to $x$ yields 
% 
\begin{equation}
    \label{eq:appx_ux_integration}
    u_{p, i}(x) 
    = 
    \sum \int_{x_0}^{x} 
    \frac{1}{u_p(x')}
    f_i(x', \tilde{y}_p(x'), \tilde{z}_p(x')) 
    \,dx'. 
\end{equation}
Note that the summation over the forcing terms has been moved outside of the integral to emphasize that contribution from each forcing term to $u_{p, i}(x)$ can be parsed separately. 
In the streamwise $i=1$ direction, \cref{eq:appx_ux_integration} can be made explicit by using the identity $2 u_p \partial u_p/\partial x = \partial (u_p)^2 / \partial x$. 
After substitution into \cref{eq:appx_ux_diff}, integrating yields
\begin{equation}
    \label{eq:appx_u1x_integration} 
    u_p(x) = \left[
    2\sum \left( 
    \int_{x_0}^x 
    f_x(x', \tilde{y}_p(x'), \tilde{z}_p(x')) \,dx'
    \right)
    + (u_p(x_0))^2
    \right]^{1/2}. 
\end{equation}

Next, we derive the integration reconstruction of the parcel position. 
In particular, we are interested in the lateral deflection $y_p(x)$ induced by $f_y(x, y, z)$. 
We apply the change of variables to \cref{eq:appx_pos_vel} in the $i=2$ direction to yield an expression for the lateral parcel velocity $u_{p, 2}(x) = v_p(x)$: 
\begin{equation}
    \label{eq:appx_vp}
    v_p(x) = u_p(x) \frac{\partial y_p}{\partial x}.
\end{equation}
Then, we substitute \cref{eq:appx_vp} into \cref{eq:appx_ux_integration} and integrate in $x$ which yields 
\begin{multline}
    \label{eq:appx_yp_terms}
    y_p(x) = 
    y_p(x_0) 
    + \underbrace{
    \int_{x_0}^x \frac{v_{p}(x_0)}{u_{p}(x')}
    \,dx' 
    }_\text{Base Advection}
    \\ + \sum 
    \underbrace{
    \int_{x_0}^{x} \frac{1}{u_{p}(x')}
    \int_{x_0}^{x'} \frac{1}{u_{p}(x'')} 
    f_y(x'', \tilde{y}_p(x''), \tilde{z}_p(x'')) \,dx'' \,dx'
    }_\text{Individual forcing contributions}. 
\end{multline}
The result in \cref{eq:appx_yp_terms} shows the decomposition of the lateral position of the fluid parcel $y_p(x)$ into the initial condition, non-linear advection from the velocity initial condition (we call this the base advection term), and individual forcing contributions. 

{
\newcommand{\fvar}{\mu}  % make it easy to change this symbol later on... 
Now we transition from streamlines to streamtubes. 
Taking a streamtube as a bundle of streamlines, we can write the streamtube-average operator for any fluid property $\fvar$ along the streamline by averaging over all of the streamlines within the streamtube. 
Using the streamtube-average notation $\langle \cdot \rangle$ from \cref{ssec:streamtubes}, we define $\langle \fvar \rangle = \langle \fvar(\tau; a_j) \rangle \forall a_j \in \Omega$, where $\Omega$ is the domain containing all streamline seed locations within the streamtube. 
Throughout this paper, we use $\Omega = \{(x, y, z) 
 | x=0, (y^2 + (z-z_h)^2) < r_s^2\}$ to seed wake streamtubes at the rotor plane with radius $r_s = 0.4D$. 
Recall that streamtube-averaging occurs over the $yz$-plane at a constant location $x$ within the streamtube. 
Then $\langle \fvar \rangle$ is a function of $x$ only, while $\fvar = \fvar(\tau; a_j) = \fvar(x, y_p(x), z_p(x))$ is a function of all three spatial coordinates. 
We define the spatial heterogeneity in the $yz$-plane by decomposing the streamtube-averaged and deviatoric components such that within the streamtube, $\fvar = \langle \fvar \rangle + \fvar''$. 
Again, $\fvar''$ is a function of all three spatial coordinates, and is only valid within the streamtube. 

Using the streamtube decomposition, we extend the equations derived for streamline integration reconstruction from the momentum budget equations to a streamtube. 
We begin with $\fvar = u_i$ to compute the streamtube-averaged velocity field. 
Streamtube-averaging \cref{eq:appx_ux_diff} over all streamlines $a_j \in \Omega$, we obtain 
\begin{equation}
    \label{eq:appx_ux_avg_diff}
    \left\langle 
    u_p(x, y_p(x), z_p(x)) \frac{\partial u_{p, i}}{\partial x} 
    \right\rangle
    = \sum 
    \langle f_i(x, \tilde{y}_p(x), \tilde{z}_p(x)) \rangle. 
\end{equation}
Spatial averaging over the RHS forcing terms can be brought inside the summation, but the advective term on the left hand side of \cref{eq:appx_ux_avg_diff} is non-linear product of two velocities. 
Again, the goal of this derivation is to connect the time-averaged wake observations, such as the wake velocity deficit and wake deflection, to the time- and streamtube-averaged budgets.
If we focused on the deflection of an individual streamline, spatial heterogeneity within the streamtube would not be relevant.
But here, we seek to understand properties for the three-dimensional wake as a whole, rather than individual streamlines.
As such, we seek to write an expression for the streamtube-averaged velocity only using streamtube-averaged quantities (such as forcing terms). 
To accomplish this, we apply the streamtube decomposition to the non-linear term in \cref{eq:appx_ux_avg_diff} such that
\begin{align}  % sort out notation stuff
    \left\langle 
    u(x, y, z) \frac{\partial u_{i}}{\partial x} 
    \right\rangle
    &=
    \left\langle 
    \big(\langle u(x, y, z) \rangle + u''(x, y, z)\big) 
    \frac{\partial}{\partial x} 
    \big(\langle u_i(x, y, z) \rangle + u_i''(x, y, z)\big)
    \right\rangle \nonumber \\
    &= 
    \label{eq:appx_dispersive}
    \langle u(x, y, z) \rangle \frac{\partial \langle u_i \rangle}{\partial x}
    + \left\langle
    u''(x, y, z) \frac{\partial u_i''}{\partial x}
    \right\rangle, 
\end{align}
where we switch the notation of $u_{p, i}(x, y_p(x), z_p(x)) \forall a_j \in \Omega$ to the Eulerian field $u_i(x, y, z)$ to denote field variables within the streamtube. 
The second RHS term in \cref{eq:appx_dispersive} is the streamtube-averaged dispersive advection. 
The dispersive advection accounts for the non-uniformity in streamwise velocity $u$ across the $yz$-cross section within the streamtube. 
% Additionally, we switch the notation of $u_{p, i}(x, y_p(x), z_p(x)) \forall a_j \in \Omega$ to the Eulerian field $u_i(x, y, z)$ to denote field variables within the streamtube. 
When the streamtube is small (see \cref{appx:streamtube_sensitivity}), dispersion can be ignored and the streamtube-averaged non-linear advection term can be approximated 
\begin{equation}
    \label{eq:appx_approx_adv}
    \left\langle u(x, y, z) \frac{\partial u_{i}}{\partial x} 
    \right\rangle
    \approx 
    \langle u(x, y, z) \rangle \frac{\partial \langle u_i \rangle}{\partial x}. 
\end{equation}
As a result, the streamtube-averaged streamwise velocity, which follows the same derivation as the streamline velocity, can be reconstructed from the RHS forcing terms as
\begin{equation}
    \label{eq:appx_u_streamtube}
    \langle u \rangle(x)
    = \left[
    2\sum \left( 
    \int_{x_0}^x 
    \langle f_x(x', y, z) \rangle \,dx'
    \right)
    + (\langle u\rangle(x_0))^2
    \right]^{1/2}. 
\end{equation}
We emphasize that the streamtube-averaged streamwise RHS forcing terms from \cref{eq:U_RANS} appear on the RHS of \cref{eq:appx_u_streamtube}, which are also the averaged forces shown in \cref{fig:streamtube_xmom}. 
}

Finally, we derive an expression for the streamtube centroid from the lateral RHS forcing terms, which is the average of the streamline positions $y_p$ within the streamtube. 
We streamtube-average over \cref{eq:appx_vp} to obtain a relation between the centroid position and the average lateral velocity: 
\begin{align}
    \label{eq:appx_vp_streamtube}
    \langle v \rangle 
    &= 
    \left\langle u(x, y, z) \frac{\partial y_p}{\partial x}
    \right\rangle \nonumber \\
    &= 
    \left\langle 
    \big(\langle u(x, y, z) \rangle + u''(x, y, z)\big) 
    \frac{\partial}{\partial x}(\langle y_p \rangle + y_p'')
    \right\rangle \nonumber \\
    &=
    \langle u \rangle \frac{\partial \langle y_p \rangle}{\partial x} + 
    \left\langle
    u''(x, y, z) \frac{\partial y_p''}{\partial x}
    \right\rangle. 
\end{align}
Again, if the streamtube is small, then the averaged advection is much larger than the dispersion and the second RHS term in \cref{eq:appx_vp_streamtube} can be ignored. 
Then, the derivation for the streamtube centroid $y_c = \langle y_p \rangle$ follows the same steps as \cref{eq:appx_yp_terms}. 
The final result is
\begin{multline}
    \label{eq:appx_yp_streamtube}
    y_c(x) = 
    \underbrace{
    \int_{x_0}^x \frac{\langle v\rangle (x_0)}{\langle u \rangle (x')}
    \,dx' 
    }_\text{Base Advection}
    + \sum 
    \underbrace{
    \int_{x_0}^{x} \frac{1}{\langle u \rangle (x')}
    \int_{x_0}^{x'} \frac{1}{\langle u \rangle (x'')} 
    \langle f_y(x'', y, z) \rangle \,dx'' \,dx'
    }_\text{Individual forcing contributions}, 
\end{multline}
where the initial wake centroid position $y_c(x_0)$ is zero in all cases because the centroid of $\Omega$ is centered about the rotor. 
The forcing terms $f_y$ are the RHS terms in the lateral RANS equation (terms II-V in \cref{eq:V_RANS}). 
Because the forcing terms $f_y$ are assumed to be steady, \cref{eq:appx_yp_streamtube} is equivalent for time-averaged flow fields $\bar{u}_i$, which is shown in \cref{eq:yc_integration}.

\end{appendix}

% ====== REFERENCES =======

\bibliographystyle{jfm}
\bibliography{references}

\end{document}